\documentclass{elsart}
\renewcommand{\theequation}{\arabic{section}.\arabic{equation}}

\newcommand{\vek}[1]{\mbox{\boldmath$#1$}}

\newcommand{\nablaleftright}{\stackrel{\leftrightarrow}{\nabla}}

\newcommand{\difleftright}{\stackrel{\leftrightarrow}{\partial}}
\newcommand{\difleft}{\stackrel{\leftarrow}{\partial}}
\newcommand{\difright}{\stackrel{\rightarrow}{\partial}}

\def\trans{\mbox{\tiny$\bot$}} % Transverse component
\def\longi{\mbox{\tiny$\|$}}   % Longitudinal component

\newcommand{\opdm}{\rho}

\begin{document}

\begin{frontmatter}
\title{Kinetic theory of QED plasmas in a strong electromagnetic field\\
I. The covariant hyperplane formalism}

\author[Rostock]{A. H\"oll\thanksref{Arne}}
\author[Moscow]{V.G. Morozov\thanksref{Vladimir}}
\author[Rostock]{G. R\"opke\thanksref{Gerd}}
\address[Rostock]{Physics Department, University of Rostock,
Universit\"atsplatz 3, D-18051 Rostock, Germany}
\address[Moscow]{Moscow State Institute of Radioengineering,
Electronics, and
Automation, 117454 Vernadsky Prospect 78, Moscow, Russia}
\thanks[Arne]{hoell@darss.mpg.uni-rostock.de}
\thanks[Vladimir]{vmorozov@orc.ru}
\thanks[Gerd]{gerd@darss.mpg.uni-rostock.de}

\begin{abstract}
We present a covariant density matrix approach to kinetic theory of QED
plasmas subjected to a strong external electromagnetic field. A canonical
quantization of the system on space-like hyperplanes in Minkowski space and a
covariant generalization of the Coulomb gauge is used. The condensate mode
associated with the mean electromagnetic field is separated from the photon
degrees of freedom by a time-dependent unitary transformation of both, the
dynamical variables and the nonequilibrium statistical operator. Therefore
even in the case of strong external fields a perturbative expansion in orders
of the fine structure constant for the
correlation functions as well as the statistical operator is applicable.
A general scheme for deriving kinetic
equations in the hyperplane formalism is presented.
\end{abstract}

\begin{keyword}
relativistic kinetic theory; QED plasma; hyperplane formalism
\end{keyword}
\end{frontmatter}
%%%%%%%%%%%%%%%%%%%%%%%%%%%%%%%%%%%%%%%%%%%%%%%%%%%%%%%%%%%%%%%%%%%%%%%%%%%%%%

\setcounter{equation}{0}
%1
\section{Introduction}
In recent years the theoretical study of dense relativistic plasmas is of
increasing interest. Such plasmas are not only limited to astrophysics, but
can nowadays be produced by high-intense short-pulse
lasers~\cite{Sprangle90,Gibbon1_96,Brabec1_00}. In view of the inertial confinement
fusion, one has to consider a plasma under extreme conditions which is
created by a strong external field. This new experimental progress needs
a systematic approach
based on quantum electrodynamics and
methods of nonequilibrium statistical mechanics.

Considerable  attention has been focussed on a mean-field (Vlasov-type)
kinetic equation for the fermionic Wigner function, which is an essential
step towards transport theory of laser-induced QED plasmas. Using the Wigner
operator defined in four-dimensional momentum
space~\cite{DeGroot80,CarruthersZachariasen83,VasakGyulassyElze87}, a
manifestly covariant mean-field kinetic equation can be derived from the
Heisenberg equations of motion for the field operators. In this approach,
however, it is difficult to formulate an initial value problem for the
kinetic equation since the four-dimensional Fourier transformation in the
covariant Wigner function includes integration of two-point correlation
functions over time. This difficulty does not appear in the scheme based on
the {\em one-time\/} fermionic Wigner function where the field operators are
taken at the same time and only the spatial Fourier transformation is
performed. In the context of QED, the one-time formulation was proposed by
Bialynicki-Birula et al.~\cite{BGR91} (referred to in the following as BGR)
and used successfully in their study of the electron-positron vacuum. Within
this approach one can explore a number of attractive features.
The one-time Wigner function has a
direct physical interpretation and allows to calculate local observables,
such as the charge density and the current density. The description  in terms
of one-time quantities is quite natural in kinetic theory based on the von
Neumann equation for the statistical operator and provides a consistent
account of causality in collision integrals.

It should
be noted, however, that the one-time Wigner function does not contain
complete information about one-particle dynamics; the spectral
properties of correlation functions can be described only in terms of
two-point Green's functions which are closely related to the covariant Wigner
function. Recently this aspect of relativistic kinetic theory was
studied within the  mean-field
approximation~\cite{ZhuangHeinz98,OchsHeinz98}. The aforementioned
incompleteness of the one-time description is well known in non-relativistic
kinetic theory, where two-time correlation functions can, in principle, be
reconstructed from the one-time Wigner function by solving integral
equations which follow from the Dyson equation for nonequilibrium Green's
functions~\cite{Lipavsky86}. The reconstruction problem in relativistic
kinetic theory remains to be explored. The solution of this problem requires
a further development of the relativistic density matrix method as well as the
relativistic Green's function technique.

In this and subsequent works we develop a density matrix
approach to kinetic theory of QED plasmas subjected to a
strong electromagnetic field. From the conceptual point of
view, our aim is to generalize the BGR scheme~\cite{BGR91} in two aspects.
First, we wish to present the one-time formalism in {\em covariant\/}
form. This removes a drawback of the BGR theory which is
not manifestly covariant. Second, we will develop a scheme which
allows to go beyond the mean-field approximation, including dissipative
processes in QED plasmas and the interplay between collisions and the
mean-field effects. Whereas subsequent studies will be concerned with
explicit kinetic equations, the present first part considers some
general problems of the
one-time covariant approach to relativistic kinetic theory.
In comparison to QED where the main object is the
$S$-matrix constructed from vacuum averages of the field operators,
kinetic theory of QED deals with averages over a nonequilibrium
ensemble describing a many-body system. Therefore we use the
Hamiltonian formalism which is typical for the density matrix method.
In this case, however, one meets with some fundamental  problems which are
considered in this paper.
In order to make
the theory manifestly covariant, canonical quantization of the system
will be carried out in a covariant fashion. Another point is that
in the presence of a strong electromagnetic field,
perturbation expansions in the fine structure constant are not suitable.
To overcome this difficulty, we will present a procedure which allows to
separate the classical part of the electromagnetic (EM) field and the
photon degrees of freedom at any time.

The paper is organized as follows. In Section~2 we briefly sketch a scheme of
relativistic statistical mechanics in the form adapted to kinetic theory. In
our approach we use a manifest covariant Schr\"odinger picture on
space-like hyperplanes in Minkowski space. Analogous formulations of
relativistic quantum mechanics and quantum field theory can be found in
literature for various applications (see,
e.g.,~\cite{Bogoliubov51,Fleming65,Fleming66,JauchRohrlich76,vanWeert82}). In
this way, ``equal-time" correlation functions are defined with respect to the
``\,invariant time'' variable on a hyperplane. In Section~3 we perform
canonical quantization of QED on space-like hyperplanes and derive the
covariant quantum Hamiltonian. Section~4 deals with the condensate mode which
corresponds to the electromagnetic field induced by the polarization
in the system. The
condensate mode is eliminated by a time-dependent unitary transformation of
the statistical operator and dynamical variables. As a result, we obtain the
effective Hamiltonian, where the interaction of fermions with the mean
electromagnetic field is incorporated non-perturbatively at any time,
while the interaction between fermions and photons is described by a
term which can be taken into account within perturbation theory.
It is shown  how Maxwell equations for the mean  electromagnetic field are
recovered in our scheme. In Section~5 the covariant one-time
Wigner function and the photon density matrix are introduced
and a method for
deriving kinetic equations in the hyperplane formalism is outlined.
The paper is summerized with a short discussion of the results and
an outlook to further applications.

We use the system of units with $c=\hbar=1$. The signature of the metric
tensor is $(+, -, -, -)$.

\setcounter{equation}{0}
%2
\section{Nonequilibrium statistical operator in the hyperplane
formalism}
%2.1
\subsection{The relativistic von Neumann equation}
It is well known that in the special theory  of relativity a quantum state of
a system is defined by a complete set of commuting observables which can be
associated with a three-parameter space-like surface $\sigma$ in Minkowski
space. Among these surfaces three-dimensional hyperplanes are especially simple
to deal with~\cite{Bogoliubov51,Fleming65,Fleming66,JauchRohrlich76}.
Since the use of
arbitrary space-like surfaces does not lead to new physics, in what follows
we  restrict our consideration to hyperplanes. A space-like hyperplane
$\sigma\equiv \sigma^{}_{n,\tau}$ is characterized by a unit time-like normal
vector $n^{\mu}$ and a scalar parameter $\tau$ which may be interpreted as an
``\,invariant time''. The equation of the hyperplane
$\sigma^{}_{n,\tau}$ reads
 \begin{equation}
 \label{EqPlane}
 x\cdot n=\tau, \qquad
n^2=n^{\mu} n^{}_{\mu}=1.
\end{equation}
In the special Lorentz frame where $n^{\mu}=(1,0,0,0)$ and
consequently Eq.~(\ref{EqPlane}) reads $x^0=\tau$ the parameter
$\tau$ coincides with the time variable $t=x^0$.
We will refer to this special frame as the ``instant frame'', since
only here observables are measured at the same instant of time $t$.
By treating a state vector
$|\Psi[\sigma^{}_{n,\tau}]\rangle$ as a functional of $\sigma^{}_{n,\tau}$,
the covariant Schr\"odinger equation can be derived from the relation between
the state vector on the hyperplane $\sigma$ and the state vector on the
hyperplane $\sigma'=L\sigma$ which is obtained by an inhomogeneous Lorentz
transformation $L=\{a,\Lambda\}$:
 \begin{equation}
 \label{LorTrf}
 \sigma\to
\sigma'=L\sigma: \quad x\to x'=\Lambda x +a.
 \end{equation}
The relation between the state vectors is~\cite{Schweber61}
 \begin{equation}
\label{TrfStVec}
U(L)\left|\Psi[L\sigma]\right\rangle=
\left|\Psi[\sigma]\right\rangle,
 \end{equation}
 where $U(L)=U(a,\Lambda)$ is a unitary
representation of the inhomogeneous Lorentz group.  The generators of this
representation, $\hat P^{\mu}$ and $\hat M^{\mu\nu}$, are the energy-momentum
vector and the angular momentum tensor, respectively.  For our purpose, the
only  transformations of relevance are pure time-like translations which change
the value of $\tau$. Recalling the form of $U(a,\Lambda)$ for pure
translations
 \begin{equation}
 \label{Transl}
 U(a,1)=\exp\left\{i\hat P^{}_{\mu} a^{\mu}\right\}
  \end{equation}
  and introducing the notation
$\left|\Psi[\sigma^{}_{n,\tau}]\right\rangle=
\left|\Psi(n,\tau)\right\rangle$, Eq.~(\ref{TrfStVec}) can be written for an
infinitesimal time-like translation $a^{\mu}=n^{\mu}\,\delta\tau$ as
 \begin{equation}
 \label{InfTrans}
 \left|\Psi(n,\tau+\delta\tau)\right\rangle +
 i\delta\tau \left(\hat P^{}_{\mu} n^{\mu}\right)
\left|\Psi(n,\tau)\right\rangle=
\left|\Psi(n,\tau)\right\rangle,
 \end{equation}
from which we obtain the relativistic Schr\"odinger  equation
 \begin{equation}
 \label{SchrEq}
 i\frac{\partial}{\partial\tau}
 \left|\Psi(n,\tau)\right\rangle= \hat H(n) \left|\Psi(n,\tau)\right\rangle
 \end{equation}
 with the Hamiltonian on the hyperplane given by
  \begin{equation}
\label{HamOnPlane}
 \hat H(n)=\hat P^{}_{\mu} n^{\mu}.
  \end{equation}
 In the presence of a
prescribed external field, the energy-momentum vector and, consequently, the
Hamiltonian $\hat H^{\tau}(n)$ can depend explicitly on $\tau$. Combining
Eq.~(\ref{SchrEq}) with the adjoint equation for the bra-vector, one finds
that the statistical operator $\varrho(n,\tau)$ for a mixed quantum ensemble
obeys the equation
 \begin{equation}
  \label{VonNEq}
   \frac{\partial\varrho(n,\tau)}{\partial\tau}
- i\left[\varrho(n,\tau),\hat H^{\tau}(n)\right]=0,
  \end{equation}
which is analogous to
the non-relativistic von Neumann equation.

%2.2
\subsection{Schr\"odinger and Heisenberg pictures on  hyperplanes}
The evolution of a mixed ensemble on space-like hyperplanes can be
represented in different pictures. The statistical operator  in the
Heisenberg picture does not depend on the parameter
$\tau$ and is associated with some fixed hyperplane
$\sigma^{}_{n,\tau^{}_{0}}$.
Dynamical variables are represented by operators
$\hat O_{H}([\sigma^{}_{n,\tau}])$ which are functionals of the
hyperplanes.
Local operators $\hat O^{}_{H}(x)$, which depend on the
space-time point $x$, are of particular interest  in quantum field
theory.
In what follows it will be convenient to treat
such operators as functions of the parameter $\tau$.
To define this dependence, we introduce the transverse projector
with respect to the normal vector $n^{\mu}$,
 \begin{equation}
 \label{TransProj}
\Delta^{\mu}_{\ \nu}=\delta^{\mu}_{\ \nu} -n^{\mu} n^{}_{\nu},
 \end{equation}
and notice that a space-time  four-vector
$x^{\mu}$ can be represented in the form
  \begin{equation}
\label{DecompX}
x^{\mu}_{}=n^{\mu}\tau + x^{\mu}_{\trans},
\qquad
\tau=n\cdot x,
  \end{equation}
where
  \begin{equation}
\label{XTrans}
x^{\mu}_{\trans}=\Delta^{\mu}_{\ \nu}\, x^{\nu}
  \end{equation}
is the transverse (space-like) component of $x$. Geometrically,
Eq.~(\ref{DecompX}) means that the space-like vector
$x^{\mu}_{\trans}$ lies on the hyperplane $\sigma^{}_{n,\tau}$
passing through the space-time point $x$.
Using the decomposition~(\ref{DecompX}), a local Heisenberg operator
$\hat O^{}_{H}(x)$ can be written as
  \begin{equation}
\label{HeisOp}
\hat O^{}_{H}(x)=\hat O^{}_{H}(n\tau +x^{}_{\trans})
\equiv
\hat O^{}_{H}(\tau, x^{}_{\trans}).
  \end{equation}
Let us assume that $\hat P^{\mu}$ does not depend explicitly on
$\tau$. Then, recalling the well-known equation of motion for
Heisenberg operators
\begin{equation}
\label{HeisEqMot}
\partial^{}_{\mu}\,\hat O^{}_{H}(x)=
-i\big[\hat O^{}_{H}(x),\hat P^{}_{\mu} \big]\, ,
\end{equation}
one readily finds that the time-like evolution of such operators is
described by the equation
\begin{equation}
\label{HeisEvol1}
\hat O^{}_{H}(\tau, x^{}_{\trans})=
{\rm e}^{i(\tau-\tau^{}_{0})\hat H(n)}\,
\hat O^{}_{H}(\tau^{}_{0}, x^{}_{\trans})\,
{\rm e}^{-i(\tau-\tau^{}_{0})\hat H(n)}
\end{equation}
with the Hamiltonian~(\ref{HamOnPlane}). The
generalization of Eq.~(\ref{HeisEvol1}) to situations in which the Hamiltonian
$\hat H^{\tau}$ depends explicitly on $\tau$ is obvious.
Defining the evolution operator $U(\tau,\tau';n)$ as the ordered exponent
\begin{equation}
\label{OrdEvOp}
U(\tau,\tau';n)=T^{}_{\tau}\,\exp\left\{
-i\int_{\tau'}^{\tau} \hat H^{\bar\tau}(n) \,d\bar\tau
\right\},
\end{equation}
we have
\begin{equation}
\label{HeisEvol2}
\hat O^{}_{H}(\tau, x^{}_{\trans})=
U^{\dagger}(\tau,\tau^{}_{0};n)\,
\hat O^{}_{H}(\tau^{}_{0}, x^{}_{\trans})\,
U^{}(\tau,\tau^{}_{0};n).
\end{equation}
In the Schr\"odinger picture, the statistical
operator  $\varrho(n,\tau)$ is $\tau$-dependent
and its time-like evolution is governed by Eq.~(\ref{VonNEq}), whereas
operators
$\hat O^{}_{S}$ are defined on a fixed hyperplane.
Assuming the Heisenberg
and Schr\"odinger pictures to coincide on the hyperplane
$\sigma^{}_{n,\tau^{}_{0}}$,  Eq.~(\ref{HeisEvol2})
implies that the transition
from the Schr\"odinger picture to the Heisenberg picture is given by
\begin{equation}
\label{FromStoH}
\hat O^{}_{H}(\tau, x^{}_{\trans})=
U^{\dagger}(\tau,\tau^{}_{0};n)\,
\hat O^{}_{S}(x^{}_{\trans})\,
U^{}(\tau,\tau^{}_{0};n).
\end{equation}
The mean values $O(x)$ of local
dynamical variables can be calculated
in both pictures. Using a formal solution of
Eq.~(\ref{VonNEq})
\begin{equation}
\label{SolVonNEq}
\varrho(n,\tau)=U(\tau,\tau^{}_{0};n)\,\varrho(n,\tau^{}_{0})\,
U^{\dagger}(\tau,\tau^{}_{0};n),
\end{equation}
we find that
\begin{equation}
\label{LocObs}
O(x)= \big\langle \hat O^{}_{H}(\tau,x^{}_{\trans})
\big\rangle^{\tau^{}_0}= \big\langle \hat O^{}_{S}(x^{}_{\trans})
\big\rangle^{\tau}.
\end{equation}
Here and in what follows the symbol $\langle\cdots\rangle^{\tau}$
stands for averages calculated with the statistical operator
$\varrho(n,\tau)$. In many problems one is dealing with partial
derivatives $\partial^{}_{\mu}O(x)$ which enter the equations of motion
for local observables.
In the hyperplane formalism, it is convenient to
express the partial derivatives in terms of the derivatives with
respect to $\tau$ and
$x^{}_{\trans}$. Recalling Eqs.~(\ref{DecompX}) and~(\ref{XTrans}), we
write
\begin{equation}
\label{DecompDer}
\partial^{}_{\mu}=
n^{}_{\mu}\frac{\partial}{\partial\tau}
+ \nabla^{}_{\mu},
\qquad
\nabla^{}_{\mu}= \Delta^{\ \nu}_{\mu} \partial^{}_{\nu}=
\Delta^{\ \nu}_{\mu}\,\frac{\partial}{\partial x^{\nu}_{\trans}}\, .
\end{equation}
Then, in the Heisenberg picture,
Eq.~(\ref{LocObs}) yields the  equation of motion
\begin{equation}
\label{ObsH:EqMot}
\partial^{}_{\mu} O(x)=
\left\langle \partial^{}_{\mu} \hat{O}^{}_{H}
(\tau,x^{}_{\trans})\right\rangle^{\tau^{}_{0}},
\end{equation}
where
\begin{eqnarray}
\label{HeisOp:EqM}
\partial^{}_{\mu} \hat{O}^{}_{H}(x)
&=&
n^{}_{\mu}\frac{\partial}{\partial\tau}\,\hat{O}^{}_{H}(\tau,x^{}_{\trans})
+\nabla^{}_{\mu} \hat{O}^{}_{H}(\tau,x^{}_{\trans})
\nonumber\\[6pt]
{}&\equiv& -in^{}_{\mu}
\left[\hat{O}^{}_{H}(\tau,x^{}_{\trans}),\hat{H}^{}_{H}(n,\tau)\right]
+\nabla^{}_{\mu} \hat{O}^{}_{H}(\tau,x^{}_{\trans}).
\end{eqnarray}
In the Schr\"odinger picture, the $\tau$-dependence of the mean values
appear through  the statistical operator which obeys the von Neumann
equation~(\ref{VonNEq}).
In this picture the equation of motion, bearing a formal
resemblance to Eq.~(\ref{ObsH:EqMot}), is obtained from
Eq.~(\ref{LocObs})\begin{equation}
\label{ObsS:EqMot}
\partial^{}_{\mu} O(x)=
\left\langle \partial^{}_{\mu} \hat{O}^{}_{S}
(x^{}_{\trans})\right\rangle^{\tau}
\end{equation}
with the analogous definition of the operator
$\partial^{}_{\mu}$ acting on local dynamical variables:
\begin{equation}
\label{SchrOp:EqM}
\partial^{}_{\mu} \hat{O}^{}_{S}(x^{}_{\trans})
=-in^{}_{\mu}
\left[\hat{O}^{}_{S}(x^{}_{\trans}),\hat{H}(n)\right]
+\nabla^{}_{\mu} \hat{O}^{}_{S}(x^{}_{\trans}).
\end{equation}

%2.3
\subsection{``Equal-time'' correlation functions}
Describing the evolution of the system  in terms of
hyperplanes, we can
introduce  ``equal-time'' correlation functions of local
dynamical variables  with respect to the invariant time $\tau$.
Let $\hat O^{}_{1H}(x), \hat O^{}_{2H}(x),\ldots,
\hat O^{}_{kH}(x)$ be some local Heisenberg operators. Then the
``equal-time"  correlation function for these operators can be
defined as
\begin{equation}
\label{SinglTCF1}
F^{}_{1\cdots k}(x^{}_{1\trans},\ldots, x^{}_{k\trans};n,\tau)=
\big\langle
\hat O^{}_{1H}(x^{}_{1})\cdots
\hat O^{}_{kH}(x^{}_{k})
\big\rangle^{\tau^{}_{0}},
\end{equation}
where $n\cdot x^{}_{1}=n\cdot x^{}_{2}=\ldots =n\cdot x^{}_{k}=\tau$.
In the Schr\"odinger picture this correlation function takes the form
\begin{equation}
\label{SinglTCF2}
F^{}_{1\cdots k}(x^{}_{1\trans},\ldots, x^{}_{k\trans};n,\tau)=
\big\langle
\hat O^{}_{1S}(x^{}_{1\trans})\cdots
\hat O^{}_{kS}(x^{}_{k\trans})
\big\rangle^{\tau}.
\end{equation}
The covariant von Neumann equation~(\ref{VonNEq}) yields the equations
\begin{equation}
\label{CFEqMot}
\hspace*{-20pt}
\frac{\partial}{\partial\tau}
F^{}_{1\cdots k}(x^{}_{1\trans},\ldots, x^{}_{k\trans};n,\tau)=
-i\big\langle
\big[
\hat O^{}_{1S}(x^{}_{1\trans})\cdots
\hat O^{}_{kS}(x^{}_{k\trans}),\hat H^{\tau}(n)
\big]
\big\rangle^{\tau}
\end{equation}
which can serve as a starting point for constructing the quantum
hierarchy for the ``equal-time" correlation functions.

\setcounter{equation}{0}
%3
\section{Hamiltonian of QED on hyperplanes}
We will now apply the foregoing scheme to  a relativistic system of charged
fermions interacting through the EM field.
For definiteness, we take these
fermions to be electrons and positrons, so that protons will be treated as a
positively charged background which ensures electric neutrality of the
system. There is no difficulty in describing protons by an additional
Dirac field.
Having in mind applications to relativistic plasmas produced by
high-intense short-pulse lasers, we assume the system to be subjected into a
prescribed external EM field which is not necessarily weak.

\subsection{The Lagrangian density}
The first step in formulating the kinetic theory of QED plasmas is to
construct the Hamiltonian $\hat H(n)$.
We start with the classical Lagrange density
\begin{equation}
\label{QED:Lagr}
{\mathcal L}(x)=
 {\mathcal L}^{}_{D}(x) + {\mathcal L}^{}_{EM}(x) +
{\mathcal L}^{}_{\rm  int}(x) +
{\mathcal L}^{}_{\rm  ext}(x),
\end{equation}
where  ${\mathcal L}^{}_{D}(x)$ and ${\mathcal L}^{}_{EM}(x)$
are the Lagrangian densities of  free Dirac  and EM fields respectively,
${\mathcal L}^{}_{\rm int}(x)$ is the interaction Lagrangian density,
and the term ${\mathcal L}^{}_{\rm  ext}(x)$ describes the interaction
of fermions with the external electromagnetic field. In
standard notation
(see, e.g.,~\cite{Greiner96}), we have
\begin{eqnarray}
\label{DirLagr}
& &
{\mathcal L}^{}_{D}(x)=
 \bar\psi(x)\left({i\over2} \gamma^{\mu}{\difleftright}^{}_{\mu}
-m \right)\psi(x),
\\[6pt]
& &
\label{EMLagr}
{\mathcal L}^{}_{EM}(x)=
-\frac{1}{4}F_{\mu\nu }(x) F^{\mu\nu }(x),
\\[6pt]
& &
\label{IntLagr}
{\mathcal L}^{}_{\rm int}(x)=
- j^{}_{\mu}(x) A^{\mu}(x),
\\[6pt]
& &
\label{ExtLagr}
{\mathcal L}^{}_{\rm ext}(x)=
- j^{}_{\mu}(x)A^{\mu}_{\rm ext}(x),
\end{eqnarray}
where
${\difleftright}^{}_{\mu}={\difright}^{}_{\mu}-{\difleft}^{}_{\mu}$.
In the following the electromagnetic field tensor is taken
in the form
$F^{}_{\mu\nu}=\partial^{}_{\mu} A^{}_{\nu}-\partial^{}_{\nu} A^{}_{\mu}$.
The current density four-vector will be expressed as
$j^{\mu}=e\bar\psi\gamma^{\mu}\psi$ with $e<0$.
We wish to remark that in our approach
the four-potential of the EM field is decomposed into two
terms. The  variables $A^{\mu}(x)$ correspond to the EM field
caused by charges and currents in the system, while
$A^{\mu}_{\rm ext}(x)$ is a prescribed external field.
In what follows, only the dynamical field $A^{\mu}(x)$ will be
quantized.

\subsection{Canonical quantization on hyperplanes}
A canonical quantization implies that some gauge fixing condition is imposed
on $A^{\mu}_{}$. For many-particle systems studied in statistical mechanics,
the Coulomb gauge seems to be the most natural. However, the disadvantage of
this gauge is that it is not manifestly covariant. Therefore we will use
a generalization of
the Coulomb gauge condition which  is consistent with the covariant description
of evolution in terms of space-like hyperplanes. To formulate this condition,
we introduce for any four-vector $V^{\mu}$ the decomposition into  the
transverse and longitudinal parts  by
 \begin{equation}
 \label{DecompVec}
 V^{\mu}=n^{\mu} V^{}_{\longi} + V^{\mu}_{\trans},
 \qquad V^{}_{\longi}=n^{}_{\nu}
 V^{\nu}, \quad V^{\mu}_{\trans}=\Delta^{\mu}_{\  \nu} V^{\nu},
 \end{equation}
where $\Delta^{\mu}_{\ \nu}$ is the projector~(\ref{TransProj}). Then a
natural generalization of the Coulomb gauge condition reads
  \begin{equation}
   \label{GaugeCond}
\nabla^{}_{\mu} A^{\mu}_{\trans}=0.
  \end{equation}
In the special frame where $n^{\mu}=(1,0,0,0)$ and $A^{\mu}=(A^0,\vek{A})$,
Eq.~(\ref{GaugeCond}) reduces to
 $\vek{\nabla}\cdot\vek{A}=0$, which is
the usual Coulomb gauge condition.

To define canonical variables for the electromagnetic field on a
hyperplane $\sigma^{}_{n,\tau}$, we first perform the
decomposition~(\ref{DecompVec}) of the field variables $A^{\mu}$ and the
decomposition~(\ref{DecompDer}) of the derivatives in the Euler-Lagrange
equations
  \begin{equation}
  \label{EulerLagr}
   \frac{\partial{\mathcal L}}{\partial A^{\mu}}
- \partial^{}_{\nu}
 \frac{\partial{\mathcal L}}{\partial(\partial^{}_{\nu} A^{\mu})}=0.
  \end{equation}
A simple algebra shows that these equations are equivalent to
  \begin{eqnarray}
  \label{EuLagrTrans1}
  & & \frac{\partial {\mathcal L}}{\partial A^{}_{\longi}}
  -\frac{\partial}{\partial\tau}
  \left(\frac{\partial{\mathcal L}}{\partial\dot{A}^{}_{\longi}}\right)
  - \nabla^{}_{\nu}\,
\frac{\partial{\mathcal L}} {\partial(\nabla^{}_{\nu} A^{}_{\longi})}=0,
\\[6pt]
 \label{EuLagrTrans2}
  & & \Delta^{\mu\nu} \left[ \frac{\partial{\mathcal L}}
 {\partial A^{\nu}_{\trans}} - \frac{\partial}{\partial\tau}
  \left(
 \frac{\partial{\mathcal L}} {\partial\dot{A}^{\nu}_{\trans} }
  \right) - \nabla^{}_{\lambda}
   \frac{\partial{\mathcal L}} {\partial(\nabla^{}_{\lambda}
  A^{\nu}_{\trans})} \right]=0,
 \end{eqnarray}
where we use the notation
 $\dot{f}\equiv\partial f/\partial\tau$
for derivatives with respect to $\tau$.
Equation~(\ref{EuLagrTrans1}) allows to eliminate the variable
$A^{}_{\longi}$ in the Lagrangian. First we rewrite
expressions~(\ref{EMLagr}) and~(\ref{IntLagr}) in terms of $A^{}_{\longi}$
and $A^{\mu}_{\trans}$ using the decomposition procedure for the derivatives
and the field $A^{\mu}$. As a result we obtain the Lagrangian density in the
form
  \begin{eqnarray}
 \label{LagrDecomp}
  & & {\mathcal L}= - {1\over4} F^{}_{\trans \mu\nu}
F^{\mu\nu}_{\trans} - {1\over2}
 \left(
\nabla^{\mu} A^{}_{\longi} - \dot{A}^{\mu}_{\trans}\right) \left(
\nabla^{}_{\mu} A^{}_{\longi} - \dot{A}^{}_{\trans\mu}\right)
\nonumber\\[6pt]
  & &
  \hspace*{90pt}
   {}- j^{}_{\longi} A^{}_{\longi} - j^{}_{\trans\mu}
  A^{\mu}_{\trans} +{\mathcal L}^{}_{D} + {\mathcal L}^{}_{\rm ext},
  \end{eqnarray}
where we have introduced the notation
  \begin{equation}
   \label{TransEMTens}
  F^{\mu\nu}_{\trans}= \nabla^{\mu} A^{\nu}_{\trans} - \nabla^{\nu}
 A^{\mu}_{\trans}.
  \end{equation}
Note that the last two terms in Eq.~(\ref{LagrDecomp}) do not contain
$A^{}_{\longi}$ and $A^{\mu}_{\trans}$. Now using the
expression~(\ref{LagrDecomp}) to calculate derivatives  in
Eq.~(\ref{EuLagrTrans1}) and taking into account that, according to the gauge
condition~(\ref{GaugeCond}), $\nabla^{}_{\mu} \dot{A}^{\mu}_{\trans}=0$, we
get
 \begin{equation}
 \label{GenPoisson} \nabla^{}_{\mu} \nabla^{\mu}
  A^{}_{\longi}= j^{}_{\longi}.
 \end{equation}
In the ``instant frame'', where
$n^{\mu}=(1,0,0,0)$, this reduces to the Poisson equation
for $A^{0}$. The solution of Eq.~(\ref{GenPoisson}) is
\begin{equation}
\label{AParall:cl}
{A}^{}_{\longi}(\tau,x^{}_{\trans})=
\int_{\sigma^{}_{n}} d\sigma'\,
G(x^{}_{\trans}-x^{\prime}_{\trans})\,{j}^{}_{\longi}
(\tau,x^{\prime}_{\trans}),
\end{equation}
where the Green function $G(x^{}_{\trans})$ satisfies the equation
\begin{equation}
\label{GrFunc:Eq}
\nabla^{}_{\mu}\nabla^{\mu} G(x^{}_{\trans})=\delta^{3}(x^{}_{\trans})
\end{equation}
with the three-dimensional delta function on a hyperplane
$\sigma^{}_{n}$ defined as
\begin{equation}
\label{DeltaFunc}
\delta^{3}(x^{}_{\trans})=
\int \frac{d^4 p}{(2\pi)^3}\, {\rm e}^{-ip\cdot x}\,
\delta(p\cdot n).
\end{equation}
The solution of Eq.~(\ref{GrFunc:Eq}) for $G(x^{}_{\trans})$ is given by
\begin{equation}
\label{GrFunc}
G(x^{}_{\trans})=
- \int \frac{d^{4}p}{(2\pi)^{3}}\, {\rm e}^{-ip\cdot x}\,
\delta(p\cdot n)\,\frac{1}{p^{2}_{\trans}}.
\end{equation}
The variable $A^{}_{\longi}$ can now be eliminated in the Lagrange
density~(\ref{LagrDecomp}) with the aid of Eq.~(\ref{AParall:cl}).
Terms like $\nabla^{}_{\nu}(\cdots)$ can be dropped
since they do not contribute to the Lagrangian
$\displaystyle L=\int {\mathcal L}\, d\sigma$
under appropriate boundary conditions. Then a straightforward  algebra
leads to
  \begin{eqnarray}
\label{LagrDecomp1}
{\mathcal L} =
&-& {1\over4} F^{}_{\trans \mu\nu} F^{\mu\nu}_{\trans}
-{1\over2} \dot{A}^{}_{\trans\mu} \dot{A}^{\mu}_{\trans}
- j^{}_{\trans\mu} A^{\mu}_{\trans}
\nonumber\\[6pt]
{}&+&
{\mathcal L}^{}_{D} + {\mathcal L}^{}_{\rm ext}
 -{1\over2}\int\limits_{\sigma^{}_{n}} d\sigma'\,
 j^{}_{\longi}(\tau,x^{}_{\trans}) G(x^{}_{\trans}-x^{\prime}_{\trans})
 j^{}_{\longi}(\tau,x^{\prime}_{\trans}).
  \end{eqnarray}
We will treat the fields $A^{\mu}_{\trans}$ as dynamical variables for
the EM field and follow
the Dirac version of  canonical quantization of
theories with constraints~\cite{Dirac50,Weinberg96}.
The gauge condition~(\ref{GaugeCond}) is one of the
constraint equation in this scheme.
Another constraint equation follows  directly from the definition of
transverse four-vectors, Eq.~(\ref{DecompVec}), and reads
\begin{equation}
\label{ConstrA1}
n^{}_{\mu} A^{\mu}_{\trans}(x)=0.
\end{equation}
We now define  canonical conjugates for the field variables
$A^{\mu}_{\trans}$ by
 \begin{equation}
\label{Momenta}
\Pi^{}_{\trans\mu}= \frac{\partial{\mathcal L}}
{\partial \dot{A}^{\mu}_{\trans}}= -\dot{A}^{}_{\trans\mu}.
\end{equation}
Obviously the $\Pi$'s are not independent variables since they
satisfy the constraint equations
$\nabla^{}_{\mu} \Pi^{\mu}_{\trans}=0$, and
$n^{}_{\mu}  {\Pi}^{\mu}_{\trans}=0$.
Thus, we have four
constraints imposed on the canonical variables. Following the standard
quantization procedure~\cite{Weinberg96}, the commutation relations for the
field operators $\hat{A}^{\mu}_{\trans}$ and $\hat{\Pi}^{\mu}_{\trans}$
can be derived. As shown in Appendix~A, these commutation relations are
\begin{eqnarray}
\label{Comm:Can}
& &
\left[\hat{A}^{\mu}_{\trans}(\tau, x^{}_{\trans}),
\hat{\Pi}^{\nu}_{\trans}(\tau, x^{\prime}_{\trans})\right]=
ic^{\mu\nu}(x^{}_{\trans} - x^{\prime}_{\trans}),
\\[5pt]
\label{Comm:CanZero}
& &
\left[\hat{A}^{\mu}_{\trans}(\tau,x^{}_{\trans}),
\hat{A}^{\nu}_{\trans}(\tau,x^{\prime}_{\trans})
\right]=\left[\hat{\Pi}^{\mu}_{\trans}(\tau,x^{}_{\trans}),
\hat{\Pi}^{\nu}_{\trans}(\tau,x^{\prime}_{\trans})
\right]=0,
\end{eqnarray}
where
\begin{equation}
\label{DiracDelta}
c^{\mu\nu}(x^{}_{\trans} -x^{\prime}_{\trans}) =
\int \frac{d^{4}p}{(2\pi)^{3}}\, {\rm e}^{-ip\cdot(x-x')}\,
\delta(p\cdot n)
\left[
\Delta^{\mu\nu}-
\frac{p^{\mu}_{\trans} p^{\nu}_{\trans}}{p^{2}_{\trans}}
\right].
\end{equation}
In Appendix~B the anticommutation relations for the Dirac field
operators on hyperplanes are derived. The result can be written as
\begin{eqnarray}
\label{Anticomm1}
& &
\bigg\{
\hat{\psi}_{a}^{} (\tau ,x_{\trans}^{}),
\,\hat{\!\bar\psi}_{\!a'} (\tau ,x_{\trans}^{\prime})
\bigg\}
=\left[\gamma^{}_{\longi}(n)\right]^{}_{aa'}
\delta_{}^{3}(x_{\trans}^{} - x_{\trans}^{\prime}),
\\[6pt]
\label{Anticomm2}
& &
\bigg\{
\hat{\psi}_{a}^{} (\tau ,x_{\trans}^{}),
\,\hat{\psi}^{}_{a'} (\tau ,x_{\trans}^{\prime})
\bigg\}
= \bigg\{
\hat{\!\bar\psi}_{\!a}^{} (\tau ,x_{\trans}^{}),
\,\hat{\!\bar\psi}_{\!a'} (\tau ,x_{\trans}^{\prime})
\bigg\}
=0,
\end{eqnarray}
where $a,\,a'$ are the spinor indices.
The matrix $\gamma^{}_{\longi}(n)$
is introduced through the following
decomposition of the Dirac matrices $\gamma^{\mu}$:
\begin{equation}
\hspace*{-25pt}
\label{Gammas:n}
\gamma^{\mu}= n^{\mu} \gamma^{}_{\longi}(n) +
\gamma^{\mu}_{\trans}(n),
\qquad
\gamma^{}_{\longi}(n)=n^{}_{\nu} \gamma^{\nu},
\quad
\gamma^{\mu}_{\trans}(n)=\left(\delta^{\mu}_{\ \nu}
- n^{\mu}  n^{}_{\nu} \right)\gamma^{\nu}.
\end{equation}
In the special Lorentz frame where
$x^{\mu}=(t,\vek{r})$ and $n^{\mu}=(1,0,0,0)$,
we have $\gamma^{}_{\longi}=\gamma^{0}$ and
$\delta_{}^{3}(x_{\trans}^{} - x_{\trans}^{\prime})=\delta(\vek{r}-\vek{r}')$,
so that
Eq.~(\ref{Anticomm1}) reduces to the well-known anticommutation
relation for the quantized Dirac field.

\subsection{Derivation of  the Hamiltonian}
The classical Hamiltonian on the hyperplane
$\sigma^{}_{n,\tau}$ can be derived in two ways. Following the
canonical procedure, $H^{\tau}(n)$ is obtained by the Legendre
transformation
\begin{equation}
\label{LegendTrans}
H(n) =
\int_{\sigma_{n,\tau}^{}}^{}d\sigma \, \bigg\{
\Pi^{}_{\trans\mu}\dot{A}_{\trans}^{\mu}
+ \bar\pi \dot{\psi}
+ \,\dot{\!\bar\psi} \pi
- \mathcal{L}^{} \bigg\} ,
\end{equation}
where ${\mathcal L}$ is given by Eq.~(\ref{LagrDecomp1}).
To find explicit expressions for the variables $\pi$ and
$\bar\pi$, which are  conjugates to the fields
$\bar\psi$ and $\psi$,
we rewrite the  Dirac Lagrangian
density~(\ref{DirLagr}) using the decomposition~(\ref{DecompDer})
of derivatives:
\begin{equation}
\label{DirLagrN}
{\mathcal L}_{D}^{} = \bar\psi
\left[
{i\over2} \left( \gamma_{\longi}^{} \frac{\difleftright}{\partial \tau}
+ \gamma_{\trans}^{\mu} \nablaleftright^{}_{\mu}\right) - m
\right]\psi.
\end{equation}
Then we have
\begin{equation}
\label{Fermi:CanMom}
\bar\pi_{}^{}\equiv
\frac{\partial{\mathcal L}_{D}^{}}{\partial \dot{\psi}}=
\frac{i}{2} \bar\psi\gamma_{\longi}^{},
\qquad
\pi_{}^{}\equiv
\frac{\partial{\mathcal L}_{D}^{}}
{\partial \,\dot{\!\bar\psi}} =
- \frac{i}{2} \gamma_{\longi}^{} \psi .
\end{equation}
Substituting expressions~(\ref{DirLagrN}) and~(\ref{Fermi:CanMom})
into Eq.~(\ref{LegendTrans}), we arrive at the classical
Hamiltonian. Another way is to start from the classical analog
of Eq.~(\ref{HamOnPlane}) which reads
\begin{equation}
\label{HamOnPlane:cl}
H(n)=P^{}_{\mu}n^{\mu}\equiv
\int_{\sigma^{}_{n,\tau}} d\sigma\,
n^{\mu}T^{}_{\mu\nu}n^{\nu},
\end{equation}
where $T^{}_{\mu\nu}(x)$ is the energy-momentum tensor.
In order that the quantized Hamiltonian  be hermitian, the
energy-momentum tensor must be real.
For instance, one
can use the so-called Belinfante tensor~\cite{Belinfante39}. When
applied to the Lagrangian~(\ref{QED:Lagr}),
the standard derivation of the Belinfante tensor
(see, e.g.,~\cite{Greiner96}) gives
\begin{eqnarray}
\label{Belinfante}
\hspace*{-10pt}
T^{}_{\mu\nu}(x)=
&-&g^{}_{\mu\nu}
\left\{
\bar\psi\left({i\over2}
\gamma^{\lambda} \difleftright^{}_{\lambda} -m\right)\psi
-j^{}_{\lambda} \left(A^{\lambda} +A^{\lambda}_{\rm ext}\right)
-{1\over4} F^{}_{\alpha\beta} F^{\alpha\beta}\right\}
\nonumber\\[6pt]
{}&+& F^{}_{\mu\lambda} F^{\lambda}_{\ \nu}
+{i\over4} \bar\psi\left(
\gamma^{}_{\nu} \difleftright^{}_{\mu} +
\gamma^{}_{\mu} \difleftright^{}_{\nu}
\right)\psi
- {1\over2} \left( j^{}_{\nu} A^{}_{\mu} + j^{}_{\mu} A^{}_{\nu}
\right).
\end{eqnarray}
Separating the
longitudinal and transverse components with respect to the
normal vector $n^{\mu}$ and then eliminating the
$\tau$-derivatives of the fields with the aid of Eqs.~(\ref{Momenta})
and~(\ref{Fermi:CanMom}), the classical Hamiltonian on the
hyperplane is obtained from Eq.~(\ref{HamOnPlane:cl}). It can
be verified that in both cases we have the same
expression for $H^{\tau}(n)$. The final step is to replace the
canonical variables
$A^{\mu}_{\trans}, \Pi^{\mu}_{\trans}, \psi, \bar{\psi}$ by the corresponding
quantum operators. As a result, we find the Hamiltonian in the
form
 \begin{equation}
\label{TotalHam:Q} \hat H^{\tau}(n)=
\hat H^{}_{D}(n)+\hat H^{}_{EM}(n)+ \hat H^{}_{\rm int}(n) +
\hat H^{\tau}_{\rm ext}(n),
\end{equation}
where $\hat H^{}_{D}(n)$
and $\hat H^{}_{EM}(n)$ are the Hamiltonians for free fermions and  the
polarization EM field respectively, $\hat H^{}_{\rm int}(n)$ is the
interaction term, and $\hat H^{\tau}_{\rm ext}(n)$
describes the external EM
field effects. In the Schr\"odinger picture the explicit
expressions for these terms are
\begin{eqnarray}
\label{DirHam:Q}
& & \hat{H}^{}_{D}(n)= \int_{\sigma^{}_{n}} d\sigma\,
\,\hat{\!\bar \psi} \left( -\frac{i}{2} \gamma^{\mu}_{\trans}(n)
\nablaleftright^{}_{\mu} +
 m \right)\hat{\psi},
\\[6pt]
\label{EMHam:Q}
& &
\hat{H}^{}_{EM}(n)=
\int_{\sigma^{}_{n}}  d\sigma\,
\left(
\frac{1}{4} \hat{F}_{\trans \mu\nu}^{} \hat{F}_{\trans}^{\mu\nu}
- \frac{1}{2} \hat{\Pi}_{\trans \mu}^{}\hat{\Pi}_{\trans}^{\mu}
\right),
\\[6pt]
\label{IntHam:Q}
& &
\hat{H}^{}_{\rm int}(n)=
\int_{\sigma^{}_{n}}  d\sigma\,
\,\hat{\!j}^{}_{\!\trans \mu} \hat{A}_{\trans}^{\mu}
+ \frac{1}{2}
\int_{\sigma^{}_{n}} d \sigma \int_{\sigma^{}_{n}}
d \sigma_{}^{\prime} \,
\,\hat{\!j}_{\!\longi}^{}(x_{\trans}^{})
G(x_{\trans}^{} - x_{\trans}^{\prime} )
\,\hat{\!j}_{\!\longi}^{}(x_{\trans}^{\prime}),
\\[6pt]
\label{ExtHam:Q}
& &
\hat{H}^{\tau}_{\rm ext}(n)=\int_{\sigma^{}_{n}} d\sigma\,
\,\hat{\!j}^{}_{\!\mu}(x^{}_{\trans}) {A}^{\mu}_{\rm ext}(\tau, x^{}_{\trans}).
\end{eqnarray}
In the  EM field Hamiltonian~(\ref{EMHam:Q}) the tensor  operator
\begin{equation}
\label{Ftens:trans}
\hat{F}^{\mu\nu}_{\trans}=
\nabla^{\mu} \hat{A}^{\nu}_{\trans} -
\nabla^{\nu} \hat{A}^{\mu}_{\trans}
\end{equation}
contains only the transverse components of the
field operators $\hat{A}^{\mu}$ which are decomposed as
\begin{equation}
\label{OpA:tot}
\hat{A}^{\mu}= n^{\mu} \hat{A}^{}_{\longi} + \hat{A}^{\mu}_{\trans}.
\end{equation}
The longitudinal part $\hat{A}^{}_{\longi}$ has been eliminated
in the interaction Hamiltonian~(\ref{IntHam:Q}) by the equation
\begin{equation}
\label{ATrans:Eq}
\nabla^{}_{\mu} \nabla^{\mu} \hat A^{}_{\longi}
=\,\hat{\!j}^{}_{\longi}
\end{equation}
which is analogous to Eq.~(\ref{GenPoisson}).
As usual, in Eqs.~(\ref{DirHam:Q})\,--\,(\ref{ExtHam:Q})
normal ordering in operators
is implied. The self-energy contribution to
the last term in Eq.~(\ref{IntHam:Q}) is
omitted, so that the product
$:\,\hat{\!j^{}_{\!\longi}}(x^{}_{\trans})\!:\,:\! \,
\hat{\!j^{}_{\!\longi}}(x'_{\trans})\!:$
is understood. For simplicity, we have written the Hamiltonian
for the case that the fermionic subsystem is described by one
Dirac field. The generalization to a many-component case is obvious.

\setcounter{equation}{0}
%4
\section{The condensate mode of the EM field}
An essential feature of the dynamical evolution
of QED plasmas in a strong external field is that
the mean values of the canonical
operators\footnote{From this point onwards symbols $A^{\mu}_{\trans}$,
$\Pi^{\mu}_{\trans}$, $j^{\mu}$, etc. denote mean values of the corresponding
operators.},
$A^{\mu}_{\trans}=\langle \hat{A}^{\mu}_{\trans}\rangle$
and  $\Pi^{\mu}_{\trans}=\langle
\hat{\Pi}^{\mu}_{\trans}\rangle$,
just as the mean values of creation and annihilation
bosonic operators $\hat{a}$ and $\hat{a}^{\dagger}$ related
to the canonical operators by plane-wave expansions, are not zero.
Furthermore, they are macroscopic quantities
associated with the mean EM field induced by the polarization in the system.
In the language of  statistical mechanics,
the variables $A^{\mu}_{\trans}(x)$ and $\Pi^{\mu}_{\trans}(x)$
describe a macroscopic {\em condensate mode\/} of the EM field.
This  fact does not allow to apply
perturbation theory directly to the
Hamiltonian~(\ref{TotalHam:Q}) because the interaction of
fermions with the condensate mode or, what is the same, with the mean
EM field is not weak. So, we have to separate the condensate
mode from the photon degrees of freedom in the Hamiltonian and
the statistical operator.

%4.1
\subsection{The time-dependent unitary transformation}
The condensate mode is most easily isolated by
introducing  the  $\tau$-dependent unitary
transformation
\begin{equation}
\label{StatOp:Transf}
\widetilde{\varrho}(n,\tau)=
{\rm e}^{i\hat{C}(n,\tau)}\,\varrho(n,\tau)\,
{\rm e}^{-i\hat{C}(n,\tau)},
\end{equation}
where the operator $\hat C(n,\tau)$ is given by
\begin{equation}
\label{OperC}
\hat{C}(n,\tau)= \int\limits_{\sigma^{}_{n}} d\sigma\,
\left\{
A^{\mu}_{\trans}(x) \hat{\Pi}^{}_{\trans\mu}(x^{}_{\trans})
- {\Pi}^{}_{\trans\mu}(x) \hat{A}^{\mu}_{\trans}(x^{}_{\trans})
\right\}.
\end{equation}
Note that the unitary transformation~(\ref{StatOp:Transf}) does not affect
fermionic operators and has the properties
\begin{eqnarray}
\label{Shift}
& &
{\rm e}^{i\hat{C}(n,\tau)}\,\hat{A}^{\mu}_{\trans}(x^{}_{\trans})\,
{\rm e}^{-i\hat{C}(n,\tau)}=
\hat{A}^{\mu}_{\trans}(x^{}_{\trans}) + A^{\mu}_{\trans}(x),
\nonumber\\[6pt]
& &
{\rm e}^{i\hat{C}(n,\tau)}\,\hat{\Pi}^{\mu}_{\trans}(x^{}_{\trans})\,
{\rm e}^{-i\hat{C}(n,\tau)}=
\hat{\Pi}^{\mu}_{\trans}(x^{}_{\trans}) + {\Pi}^{\mu}_{\trans}(x).
\end{eqnarray}
Taking now into account that, for any operator $\hat{O}$,
\begin{equation}
\label{Avergs:Trans}
\langle \hat{O} \rangle^{\tau}=
{\rm Tr} \left\{
{\rm e}^{i\hat{C}(n,\tau)}\,\hat{O}\,
{\rm e}^{-i\hat{C}(n,\tau)}\,\widetilde{\varrho}(n,\tau)
\right\} \equiv
\left\langle {\rm e}^{i\hat{C}(n,\tau)}\,\hat{O}\,
{\rm e}^{-i\hat{C}(n,\tau)}\right\rangle^{\tau}_{\tilde{\varrho}},
\end{equation}
we find that
\begin{equation}
\label{ZeroCond}
\left\langle
\hat{A}^{\mu}_{\trans}(x^{}_{\trans})
\right\rangle^{\tau}_{\tilde{\varrho}}=
\left\langle\hat{\Pi}^{\mu}_{\trans}(x^{}_{\trans})
\right\rangle^{\tau}_{\tilde{\varrho}}=0.
\end{equation}
Thus, in the state described by the transformed
statistical operator~(\ref{StatOp:Transf}), the
canonical dynamical variables $\hat{A}^{\mu}_{\trans}$ and
$\hat{\Pi}^{\mu}_{\trans}$ have zero
mean values and, hence, they are not related to  the
condensate mode. In other words, after the unitary transformation the
EM field operators correspond to the photon degrees of freedom.
Based on the above arguments, it is convenient to use
$\widetilde{\varrho}(n,\tau)$ as the statistical operator of the system.
It should be noted, however, that $\widetilde{\varrho}(n,\tau)$ does not
satisfy the von Neumann equation~(\ref{VonNEq}) since the operator
$\hat{C}$ depends on $\tau$. In order to derive the equation of motion
for $\widetilde{\varrho}(n,\tau)$, we differentiate
Eq.~(\ref{StatOp:Transf}) with respect to $\tau$. After some algebra
which we omit, we find that the transformed statistical operator
satisfies the modified von Neumann equation
\begin{equation}
\label{VonNEq:Trnsf}
\frac{\partial\widetilde{\varrho}(n,\tau)}{\partial\tau} -
i\left[\widetilde{\varrho}(n,\tau),\hat{\mathcal H}^{\tau}(n)\right]=0
\end{equation}
with the effective Hamiltonian
\begin{eqnarray}
\label{EffHam:1}
& &
\hat{\mathcal H}^{\tau}(n)=
{\rm e}^{i\hat{C}(n,\tau)}\,\hat{H}^{\tau}(n)\,
{\rm e}^{-i\hat{C}(n,\tau)}
\nonumber\\[8pt]
& &
\hspace*{80pt}
{}+
\int\limits_{\sigma^{}_{n}} d\sigma
\left\{
\frac{\partial {\Pi}^{}_{\trans\mu}(x)}{\partial\tau} \,
\hat{A}^{\mu}_{\trans}(x^{}_{\trans})
-\frac{\partial {A}^{\mu}_{\trans}(x)}{\partial\tau} \,
\hat{\Pi}^{}_{\trans\mu}(x^{}_{\trans})
\right\}.
\end{eqnarray}
The transformation of  $\hat{H}(n)$ in the first term is trivial
due to Eqs.~(\ref{Shift}) and the fact that the transformation
does not affect fermionic operators.
It is convenient to eliminate the derivatives in the last term of
Eq.~(\ref{EffHam:1}) with the aid of the equations of
motion for the condensate mode
\begin{equation}
\label{Condens:Eqs}
\frac{\partial A^{\mu}_{\trans}(x)}{\partial\tau}=
-\Pi^{\mu}_{\trans}(x),
\qquad
\frac{\partial \Pi^{\mu}_{\trans}(x)}{\partial\tau}=
\nabla^{}_{\lambda} F^{\lambda\mu}_{\trans}(x)
-j^{\mu}_{\trans}(x),
\end{equation}
which are easily derived using
Eqs.~(\ref{ObsS:EqMot}),~(\ref{SchrOp:EqM}),
and the canonical commutation relations~(\ref{Comm:Can}).
The tensor $F^{\mu\nu}_{\trans}$ in Eq.~(\ref{Condens:Eqs})
is the mean value of the operator~(\ref{Ftens:trans}), and
$j^{\mu}_{\trans}(x)$ is the transverse part of the mean
polarization current
\begin{equation}
\label{PolCurr}
j^{\mu}(x)=\langle \,\,\hat{\!j}^{\,\mu}(x^{}_{\trans})\rangle^{\tau}.
\end{equation}
Inserting Eqs.~(\ref{Condens:Eqs}) into
Eq.~(\ref{EffHam:1}), the effective Hamiltonian
can be written as a sum
\begin{equation}
\label{EffHam:2}
\hat{\mathcal H}^{\tau}(n)= \hat{\mathcal H}^{\tau}_{0}(n)
+ \hat{\mathcal H}^{\tau}_{\rm int}(n).
\end{equation}
The main  term
\begin{equation}
\label{ZeroHam}
\hat{\mathcal H}^{\tau}_{0}(n)=
\hat{H}^{}_{D}(n) + \hat{H}^{}_{EM} +
\int\limits_{\sigma^{}_{n}} d\sigma\,
\,\hat{\!j}^{}_{\mu}(x^{}_{\trans})\,{\mathcal A}^{\mu}(x)
\end{equation}
describes  free photons and fermions interacting with the total
electromagnetic field
\begin{equation}
\label{EM:totA}
{\mathcal A}^{\mu}(x) = A^{\mu}_{\rm ext}(x) +
A^{\mu}(x),
\end{equation}
where the mean polarization field $A^{\mu}(x)$ is given by
\begin{equation}
\label{EM:polA}
A^{\mu}(x)= \left\langle \hat{A}^{\mu}(x^{}_{\trans})
\right\rangle^{\tau}.
\end{equation}
The term  $\hat{\mathcal H}^{\tau}_{\rm int}(n)$
 in Eq.~(\ref{EffHam:2}) describes a weak interaction between
photons and fermions. The explicit expression for this term is
\begin{eqnarray}
\label{EffInt}
& &
\hat{\mathcal H}^{\tau}_{\rm int}(n)=
\int\limits^{}_{\sigma^{}_{n}} d\sigma\,
\Delta\,\hat{\!j}^{\,\mu}_{\trans}(x^{}_{\trans};\tau)\,
\hat{A}^{\mu}_{\trans}(x^{}_{\trans})
\nonumber\\[6pt]
& &
\hspace*{60pt}
{}+{1\over2} \int\limits_{\sigma^{}_{n}} d\sigma
\int\limits_{\sigma^{}_{n}} d\sigma' \,
\Delta\,\hat{\!j}^{}_{\longi}(x^{}_{\trans};\tau)\,
G(x^{}_{\trans} - x^{\prime}_{\trans})\,
\Delta\,\hat{\!j}^{}_{\longi}(x^{\prime}_{\trans};\tau),
\end{eqnarray}
where the operators
\begin{equation}
\label{DeltaJ}
\Delta\,\hat{\!j}^{\,\mu}(x^{}_{\trans};\tau)=
\hat{\!j}^{\,\mu}(x^{}_{\trans})-
\langle \,\,\hat{\!j}^{\,\mu}(x^{}_{\trans}) \rangle^{\tau}
\end{equation}
represent quantum fluctuations of the fermionic current.
The essential point is that now the interaction term~(\ref{EffInt})
does not contain a contribution from the condensate mode
and, consequently, one can use
perturbation expansions in the fine structure constant.

%4.2
\subsection{Maxwell equations}
To complete our discussion of the condensate mode, we will show how
the Maxwell equations for the total  mean EM are derived in our
approach. According to Eq.~(\ref{EM:totA}), the total field tensor can be
written as
\begin{equation}
\label{MeanF:Tensor}
{\mathcal F}^{\mu\nu}(x)=
\partial^{\mu} {\mathcal A}^{\nu}(x)
-\partial^{\nu} {\mathcal A}^{\mu}(x)=
F^{\mu\nu}_{\rm ext}(x) +
F^{\mu\nu}(x).
\end{equation}
The external field tensor $F^{\mu\nu}_{\rm ext}$ is assumed to
satisfy the Maxwell equations
\begin{equation}
\label{Maxwell:Ext}
\partial^{}_{\mu} {F}^{\mu\nu}_{\rm ext}(x)=
j^{\nu}_{\rm ext}(x)
\end{equation}
with some prescribed
external current $j^{\mu}_{\rm ext}$.
On the other hand, the polarization field tensor $F^{\mu\nu}(x)$ is
the mean value  of the operator
\begin{eqnarray}
\hat{F}^{\mu\nu}(x^{}_{\trans})
&=&\partial^{\mu}\hat{A}^{\nu}-\partial^{\nu}\hat{A}^{\mu}
 \nonumber\\[6pt]
\label{Ftens:pol}
{}&=&
\hat{F}^{\mu\nu}_{\trans} +
n^{\nu}\left(\hat{\Pi}^{\mu}_{\trans} +
\nabla^{\mu} \hat{A}^{}_{\longi}\right)
- n^{\mu}\left(\hat{\Pi}^{\nu}_{\trans} +
\nabla^{\nu} \hat{A}^{}_{\longi}\right).
\end{eqnarray}
Recalling Eqs.~(\ref{ObsS:EqMot}) and~(\ref{SchrOp:EqM}),
straightforward algebraic
manipulations with the equations of motion for
the field operators lead to the Maxwell equations for the polarization
field tensor
\begin{equation}
\label{Maxwell:Pol}
\partial^{}_{\mu} F^{\mu\nu}(x)=j^{\nu}(x).
\end{equation}
Now  Eqs.~(\ref{Maxwell:Ext}) and~(\ref{Maxwell:Pol}) can be combined
into the Maxwell equations for the total  field tensor
\begin{equation}
\label{Maxwell:Tot}
\partial^{}_{\mu} {\mathcal F}^{\mu\nu}(x)=j^{\nu}(x) +
j^{\nu}_{\rm ext}(x).
\end{equation}
A solution of these equations gives the total mean  field
${\mathcal A}^{\mu}$ in terms of the
total mean current.

\setcounter{equation}{0}
%5
\section{Kinetic description of QED plasmas}
%5.1
\subsection{The ``one-time"  Wigner function}
Within the hyperplane formalism a natural way of describing kinetic
processes in the fermion subsystem is by the ``one-time"
Wigner function which depends on the
variable $\tau$.
Since there is the gauge freedom for the mean field
${\mathcal A}^{\mu}$, it is convenient to employ
the gauge-invariant  Wigner function on the hyperplane
$\sigma^{}_{n,\tau}$ defined as
\begin{eqnarray}
\label{Wigner:Def}
& &
\hspace*{-20pt}
W^{}_{aa'}(x^{}_{\trans},p^{}_{\trans};\tau)=
\int d^4 y\, {\rm e}^{ip\cdot y}\,\delta(y\cdot n)\,
\nonumber\\[6pt]
& &
%\hspace*{30pt}
{}\times
\exp\left\{
ie\Lambda(x^{}_{\trans} +\mbox{$1\over2 $}y^{}_{\trans},
x^{}_{\trans}-\mbox{$1\over2 $}y^{}_{\trans};\tau)
\right\}
\opdm^{}_{aa'}
\left(x^{}_{\trans}+\mbox{$1\over2 $}y^{}_{\trans},
x^{}_{\trans}-\mbox{$1\over2 $}y^{}_{\trans};\tau\right)
\end{eqnarray}
with the gauge function
\begin{eqnarray}
\label{GaugeFunc}
\Lambda(x^{}_{\trans},x^{\prime}_{\trans};\tau)
&=&
\int\limits_{x^{\prime}_{\trans}}^{x^{}_{\trans}}
{\mathcal A}^{}_{\trans\mu}(\tau,R^{}_{\trans})\,
dR^{\mu}_{\trans}
\nonumber\\[6pt]
{}&\equiv&
\int\limits_{0}^{1}
ds \left(x^{\mu}_{\trans} - x^{\prime\mu}_{\trans}\right)
{\mathcal A}^{}_{\trans\mu}\big(
\tau, x^{\prime}_{\trans} +s(x^{}_{\trans} -
x^{\prime}_{\trans}) \big).
\end{eqnarray}
In Eq.~(\ref{Wigner:Def}) the one-particle density
matrix $\rho^{}_{aa'}$ is the mean value
\begin{equation}
\label{DensMatr:Def}
\opdm^{}_{aa'}
\left(x^{}_{\trans}, x^{\prime}_{\trans} ;\tau\right)=
\left\langle
\hat \opdm^{}_{aa'}
\left(x^{}_{\trans}, x^{\prime}_{\trans}\right)
\right\rangle^{\tau}=
\left\langle
\hat \opdm^{}_{aa'}
\left(x^{}_{\trans}, x^{\prime}_{\trans}\right)
\right\rangle^{\tau}_{\tilde\varrho}
\end{equation}
of some density operator $\hat \opdm^{}_{aa'}$.
In the literature one can find different
definitions for the fermionic  density operator.
The most often used definitions  are
\begin{eqnarray}
\label{DensOper:Com}
& &
\hat \opdm^{}_{aa'}(x^{}_{\trans},x^{\prime}_{\trans})=
-{1\over2}\big[
\hat\psi^{}_{a}(x^{}_{\trans}),
\,\hat{\!\bar \psi}^{}_{a'}(x^{\prime}_{\trans})
\big],
\\[6pt]
\label{DensOper:N}
& &
\hat \opdm^{\prime}_{aa'}(x^{}_{\trans},x^{\prime}_{\trans})=
\,:\hat{\!\bar \psi}^{}_{a'}(x^{\prime}_{\trans})\,
\,\hat{\!\psi}^{}_{a}(x^{}_{\trans})\!:\, .
\end{eqnarray}
These two operators are related by
\begin{equation}
\label{DensOps:Rel}
\hat \opdm^{}_{aa'}(x^{}_{\trans},x^{\prime}_{\trans})=
\hat \opdm^{\prime}_{aa'}(x^{}_{\trans},x^{\prime}_{\trans})   +
K^{}_{aa'}(x^{}_{\trans},x^{\prime}_{\trans}),
\end{equation}
where the last $c$-number term represents  the vacuum
expectation value of $\hat{\opdm}$ since the vacuum expectation
value of $\hat \opdm^{\prime}$ is zero.
It can be shown, however, that  the vacuum term in
Eq.~(\ref{DensOps:Rel})
does not contribute to local observables like the mean current
$j^{\mu}(x)$.
The advantage of the definition~(\ref{DensOper:N})
is that  the mean values of one-particle dynamical
variables (summation over repeated spinor indices)
\begin{equation}
\label{OnePartOp}
\hat{O}=
\int\limits_{\sigma^{}_{n}}d\sigma\, d\sigma'\,
O^{}_{a'a}(x^{\prime}_{\trans},x^{}_{\trans})
:\hat{\!\bar \psi}^{}_{a'}(x^{\prime}_{\trans})\,
\,\hat{\!\psi}^{}_{a}(x^{}_{\trans})\!:
\end{equation}
are conveniently expressed in terms of the density matrix
$\opdm'=\langle\hat \opdm^{\prime}\rangle^{\tau}$:
\begin{equation}
\label{OnePartObs}
\langle \hat O\rangle^{\tau}=
\int\limits_{\sigma^{}_{n}}d\sigma\, d\sigma'\,
O^{}_{a'a}(x^{\prime}_{\trans},x^{}_{\trans})
\opdm^{\prime}_{aa'}(x^{}_{\trans},x^{\prime}_{\trans};\tau).
\end{equation}
Unfortunately, the equation of motion for
the density operator~(\ref{DensOper:N}) with the
Hamiltonian~(\ref{EffHam:2})
contains  vacuum (divergent) terms. On the other
hand,  such terms do not appear in the equation of motion for
the operator~(\ref{DensOper:Com}). For this reason, we
shall take the operator~(\ref{DensOper:Com}) as the
one-particle density operator in Eq.~(\ref{DensMatr:Def}).
An analogous definition
was used previously for the phase-space description of the QED
vacuum in a strong field~\cite{BGR91}.

The Wigner function~(\ref{Wigner:Def}) is defined on a given
family of hyperplanes $\sigma^{}_{n,\tau}$ and, hence, depends
parametrically on the normal four-vector $n$.
It should be noted, however, that
local observables calculated from the Wigner function
do not depend on the choice of $n$.
As an important example, the mean polarization current~(\ref{PolCurr})
can be written in the form
 \begin{eqnarray}
 \label{Wigner:Curr}
  j^{\mu}(x)&=& e\big\langle
:\hat{\!\bar\psi}(x^{}_{\trans}) \gamma^{\mu} \hat{\psi}(x^{}_{\trans})\!:
\big\rangle^{\tau}
\nonumber\\[6pt]
{}&=& e\int\frac{d^4p}{(2\pi)^3}\,\delta(p\cdot n)\,
 {\rm tr}\left\{\gamma^{\mu}
W(x^{}_{\trans},p^{}_{\trans};\tau=x\cdot n) \right\},
\end{eqnarray}
where the
symbol ``${\rm tr}$'' stands for the trace over spinor indices.
Geometrically, the above relation means that, in calculating the current, the
invariant time $\tau$ has a value such that the space-time point $x$ lies on
the hyperplane $\sigma^{}_{n,\tau}$.

\subsection{The photon density matrix}
To define the photon density matrix, we start from the plane wave
expansion of the vector potential operator $\hat{A}^{}_{\trans}$
in terms of creation and annihilation operators.
By analogy with the well-known representation for the free photon field in the
special Lorentz frame where $n^{\mu}=(1,0,0,0)$, we write
 \begin{eqnarray}
 \label{Exp:A}
& &
\hspace*{-15pt}
 \hat{A}^{\mu}_{\trans}(\tau,x^{}_{\trans})=
 \int \frac{d^4 q}{\sqrt{2\omega^{}_{n}(q^{}_{\trans})(2\pi)^{3}}}\,
 \delta\!\left(q^{}_{\longi} -\omega^{}_{n}(q^{}_{\trans})\right)
\nonumber\\[8pt]
& &
\hspace*{60pt}
{}\times
 \sum_{l=1,2} e^{\mu}(q^{}_{\trans},l) \left\{
 \hat{a}^{}_{l}(q^{}_{\trans})\,{\rm e}^{-iq\cdot x} +
 \hat{a}^{\dagger}_{l}(q^{}_{\trans})\,{\rm e}^{iq\cdot x}
 \right\},
 \end{eqnarray}
where $e^{\mu}(q^{}_{\trans},l)$ are real-valued polarization
four-vectors and
 \begin{equation}
 \label{Phot:Disp}
 \omega^{}_{n}(q^{}_{\trans})=\omega^{}_{n}(-q^{}_{\trans})=
 \left( -q^{\mu}_{\trans} q^{}_{\trans\mu}\right)^{1/2}
 \end{equation}
is the dispersion relation for free photons on the hyperplane.
The conditions
$\nabla^{}_{\mu} \hat{A}^{\mu}_{\trans}=
n^{}_{\mu} \hat{A}^{\mu}_{\trans}=0$ mean that the polarization vectors
satisfy
 \begin{equation}
 \label{PolV:Prop}
 q^{}_{\trans\mu} e^{\mu}(q^{}_{\trans},l)
 =n^{}_{\mu} e^{\mu}(q^{}_{\trans},l)=0.
 \end{equation}
The expansion of the operator $\hat{\Pi}^{\mu}_{\trans}$ into plane waves
is found from~(\ref{Exp:A}) by using
$\hat{\Pi}^{\mu}_{\trans}=-\!\dot{\,\hat{A}^{\mu}_{\trans}}$:
 \begin{eqnarray}
 \label{Exp:Pi}
& &
\hspace*{-15pt}
 \hat{\Pi}^{\mu}_{\trans}(\tau,x^{}_{\trans})=
 \int \frac{d^4 q}{\sqrt{2\omega^{}_{n}(q^{}_{\trans})(2\pi)^{3}}}\,
 i\omega^{}_{n}(q^{}_{\trans})\,
 \delta\!\left(q^{}_{\longi} -\omega^{}_{n}(q^{}_{\trans})\right)
\nonumber\\[8pt]
& &
\hspace*{60pt}
{}\times
 \sum_{l=1,2} e^{\mu}(q^{}_{\trans},l) \left\{
 \hat{a}^{}_{l}(q^{}_{\trans})\,{\rm e}^{-iq\cdot x} -
 \hat{a}^{\dagger}_{l}(q^{}_{\trans})\,{\rm e}^{iq\cdot x}
 \right\}.
 \end{eqnarray}
Assuming the commutation relations for the creation and annihilation
operators
\begin{eqnarray}
\label{Comm:Phot}
 & &
 \big[\hat{a}^{}_{l}(q^{}_{\trans}),
  \hat{a}^{\dagger}_{l'}(q^{\prime}_{\trans})\big]=
 \delta^{}_{ll'} \,\delta^{3}(q^{}_{\trans} -q^{\prime}_{\trans}),
 \nonumber\\[-4pt]
& &
 {}\qquad
 \\[-4pt]
 & &
\big[\hat{a}^{}_{l}(q^{}_{\trans}),
  \hat{a}^{}_{l'}(q^{\prime}_{\trans})\big]=
  \big[\hat{a}^{\dagger}_{l}(q^{}_{\trans}),
  \hat{a}^{\dagger}_{l'}(q^{\prime}_{\trans})\big]=0,
\nonumber
\end{eqnarray}
and the completeness relation for the polarization vectors
\begin{equation}
\label{ComplRel}
\sum_{l=1,2} e^{\mu}(q^{}_{\trans},l)\,
e^{\nu}(q^{}_{\trans},l)=
-\left(
\Delta^{\mu\nu} -\frac{q^{\mu}_{\trans}q^{\nu}_{\trans}}{q^{2}_{\trans}}
\right),
\end{equation}
the commutation relation~(\ref{Comm:Can}) for the field operators
is recovered.
Note that Eqs.~(\ref{Exp:A}) and~(\ref{Exp:Pi}) give the field operators in
the interaction picture. The corresponding expansions for the field operators
in the Schr\"odinger picture are obtained by setting $\tau=0$. In
this case the delta-function
$\delta(q^{}_{\longi}-\omega^{}_{n}(q^{}_{\trans}))$ can be replaced by
$\delta(q^{}_{\longi})$.

The above considerations suggest that it is natural to define the photon
density matrix in terms of the Schr\"odinger operators
 \begin{equation}
 \label{PhotField}
\begin{array}{l}
\displaystyle
\hat{\varphi}^{}_{l}(x^{}_{\trans})= \int \frac{d^4 q}{(2\pi)^{3/2}}\,
\delta(q^{}_{\longi})\,
{\rm e}^{-iq\cdot x} \hat{a}^{}_{l}(q^{}_{\trans}),
\\[10pt]
\displaystyle
\hat{\varphi}^{\dagger}_{l}(x^{}_{\trans})= \int \frac{d^4 q}{(2\pi)^{3/2}}\,
\delta(q^{}_{\longi})\,
{\rm e}^{iq\cdot x} \hat{a}^{\dagger}_{l}(q^{}_{\trans}),
\end{array}
 \end{equation}
which satisfy the commutation relations
 \begin{equation}
 \label{Comm:Phi}
 \begin{array}{l}
 \big[
 \hat{\varphi}^{}_{l}(x^{}_{\trans}),
 \hat{\varphi}^{\dagger}_{l'}(x^{\prime}_{\trans})\big]= \delta^{}_{ll'}\,
 \delta^{3}(x^{}_{\trans}-x^{\prime}_{\trans}),
 \\[6pt]
\big[
 \hat{\varphi}^{}_{l}(x^{}_{\trans}),
 \hat{\varphi}^{}_{l'}(x^{\prime}_{\trans}) \big]=
 \big[\hat{\varphi}^{\dagger}_{l}(x^{}_{\trans}),
 \hat{\varphi}^{\dagger}_{l'}(x^{\prime}_{\trans})\big]=0.
 \end{array}
 \end{equation}
The photon density matrix is defined as
 \begin{equation}
 \label{PhotDensM}
 N^{}_{ll'}(x^{}_{\trans},x^{\prime}_{\trans};\tau)=
 \left\langle
\hat{N}^{}_{ll'}(x^{}_{\trans},x^{\prime}_{\trans})
 \right\rangle^{\tau}_{\tilde\varrho},
\end{equation}
where
\begin{equation}
\label{PhotDensOp}
\hat{N}^{}_{ll'}(x^{}_{\trans},x^{\prime}_{\trans})=
\hat{\varphi}^{\dagger}_{l'}(x^{\prime}_{\trans})\,
\hat{\varphi}^{}_{l}(x^{}_{\trans})
 \end{equation}
is the photon density operator. It should be emphasized that in
Eq.~(\ref{PhotDensM}) the average is calculated with the
transformed statistical
operator $\widetilde{\varrho}(n,\tau)$ in which the condensate mode of EM
field has been eliminated. When written in terms of the average with
the statistical operator $\varrho(n,\tau)$,
the photon density matrix takes the form
 \begin{equation}
 \label{PhotdensM1}
 N^{}_{ll'}(x^{}_{\trans},x^{\prime}_{\trans};\tau)=
 \left\langle
\hat{N}^{}_{ll'}(x^{}_{\trans},x^{\prime}_{\trans})
 \right\rangle^{\tau}-
 \langle \hat{\varphi}^{}_{l}(x^{}_{\trans})\rangle^{\tau}\,
\langle \hat{\varphi}^{\dagger}_{l'}(x^{\prime}_{\trans})\rangle^{\tau},
 \end{equation}
where the last term corresponds to the contribution from the condensate mode.

\subsection{The covariant statistical operator in QED kinetics}
The evolution of the fermionic Wigner function~(\ref{Wigner:Def}) and
the photon density matrix~(\ref{PhotDensM}) is governed by kinetic
equations which can be derived from the equations of motions
 \begin{eqnarray}
 \label{FDens:EqMot}
 & &
 \hspace*{-20pt}
\frac{\partial}{\partial\tau}\,
\opdm^{}_{aa'}(x^{}_{\trans},x^{\prime}_{\trans};\tau)= -i\,
{\rm Tr}\left\{
 \big[
\hat \opdm^{}_{aa'}(x^{}_{\trans},x^{\prime}_{\trans}),
\hat{\mathcal H}^{\tau}_{0}(n)
+ \hat{\mathcal H}^{\tau}_{\rm int}(n)
\big]\,\widetilde{\varrho}(n,\tau)\right\},
\\[8pt]
\label{PDens:EqMot}
& &
\hspace*{-20pt}
\frac{\partial}{\partial\tau}\,
N^{}_{ll'}(x^{}_{\trans},x^{\prime}_{\trans};\tau)= -i\,
{\rm Tr}\left\{
\big[
\hat{N}^{}_{ll'} (x^{}_{\trans},x^{\prime}_{\trans}),
\hat{\mathcal H}^{\tau}_{0}(n)
+ \hat{\mathcal H}^{\tau}_{\rm int}(n)
\big]\,\widetilde{\varrho}(n,\tau)\right\}.
\end{eqnarray}
There are two ways to express the right-hand sides of
these equations in terms of the fermionic and photon density matrices using
perturbation expansions in the fine structure constant. One method
is by considering the hierarchy for correlation functions which
appear through the commutators with the interaction Hamiltonian
$\hat{\mathcal H}^{\tau}_{\rm int}(n)$ and then employing some truncation
procedure. Another method is to construct an approximate solution of
Eq.~(\ref{VonNEq:Trnsf}) in
terms of the density matrices $\rho$ and $N$. In both cases one has to
impose some boundary conditions of the retarded type on the correlation
functions or the statistical operator. The standard boundary condition in
kinetic theory is Bogoliubov's boundary condition of weakening of initial
correlations which implies the uncoupling of all correlation functions to
one-particle density matrices in the distant past, i.e., for
$\tau\to -\infty$.
In the scheme
developed by Zubarev (see, e.g.,~\cite{ZubMorRoep1}),
such boundary conditions can be included by using
instead of Eq.~(\ref{VonNEq:Trnsf})
the equation with an infinitesimally small source term
 \begin{equation}
 \label{ZubEq}
\frac{\partial\widetilde{\varrho}(n,\tau)}{\partial\tau} -
i\left[\widetilde{\varrho}(n,\tau),\hat{\mathcal H}^{\tau}(n)\right]=
-\varepsilon\left\{
\widetilde{\varrho}(n,\tau)-\varrho^{}_{\rm rel}(n,\tau)
\right\},
 \end{equation}
where $\varepsilon\to +0$ after the calculation of averages.
Here $\varrho^{}_{\rm rel}(n,\tau)$ is the so-called
\emph{relevant statistical operator\/} which describes a Gibbs state
for some given nonequilibrium state variables.
In QED kinetics these variables are
the Wigner function~(\ref{Wigner:Def}) and the photon density
matrix~(\ref{PhotDensM}). Therefore, following the standard
procedure~\cite{ZubMorRoep1}, we obtain the relevant statistical operator
in the form (with summation over spinor and polarization indices)
 \begin{eqnarray}
 \label{RelStOp}
 & &
 \hspace*{-20pt}
\varrho^{}_{\rm rel}(n,\tau)=
{1\over Z^{}_{\rm rel}(n,\tau)}\,
\exp\Bigg\{
- \int\limits_{\sigma^{}_{n}} d\sigma\,d\sigma'
\bigg[
\lambda^{(f)}_{aa'} (x^{}_{\trans},x^{\prime}_{\trans};\tau)
\,:\hat{\!\bar \psi}^{}_{a}(x^{}_{\trans})\,
\,\hat{\!\psi}^{}_{a'}(x^{\prime}_{\trans})\!:
\nonumber\\[8pt]
& &
\hspace*{140pt}
{}+\lambda^{(ph)}_{ll'} (x^{}_{\trans},x^{\prime}_{\trans};\tau)\,
\hat{\varphi}^{\dagger}_{l}(x^{}_{\trans})\,
\hat{\varphi}^{}_{l'}(x^{\prime}_{\trans})
\bigg]
\Bigg\},
 \end{eqnarray}
where $Z^{}_{\rm rel}(n,\tau)$ is the normalization constant
(or the partition function in the relevant ensemble) and
$\lambda^{(f)}_{aa'} (x^{}_{\trans},x^{\prime}_{\trans};\tau)$,
$\lambda^{(ph)}_{ll'} (x^{}_{\trans},x^{\prime}_{\trans};\tau)$ are
Lagrange multipliers which are determined by the self-consistency
conditions
 \begin{equation}
\label{SelfCons}
 \begin{array}{l}
 \displaystyle
\opdm^{}_{aa'}(x^{}_{\trans},x^{\prime}_{\trans};\tau)=
{\rm Tr}\left\{\hat{\opdm}^{}_{aa'}(x^{}_{\trans},x^{\prime}_{\trans})
\varrho^{}_{\rm rel}(n,\tau) \right\},
\\[6pt]
\displaystyle
N^{}_{ll'}(x^{}_{\trans},x^{\prime}_{\trans};\tau) =
{\rm Tr}\left\{\hat{N}^{}_{ll'}(x^{}_{\trans},x^{\prime}_{\trans})
\varrho^{}_{\rm rel}(n,\tau) \right\}.
\end{array}
 \end{equation}
Using Eq.~(\ref{ZubEq}) for the transformed statistical operator
leads to the hierarchy
\begin{eqnarray}
\label{TrCorr:Eq}
& &
\hspace*{-20pt}
\frac{\partial}{\partial\tau}
\widetilde{F}^{}_{1\cdots k}(x^{}_{1\trans},\ldots, x^{}_{k\trans};n,\tau)=
-i\left\langle
\big[
\hat O^{}_{1}(x^{}_{1\trans})\cdots
\hat O^{}_{k}(x^{}_{k\trans}),\hat{\mathcal H}^{\tau}(n)
\big]
\right\rangle^{\tau}_{\tilde{\varrho}}
\nonumber\\[8pt]
& &
\hspace*{20pt}
{}-\varepsilon
\left\{
\widetilde{F}^{}_{1\cdots k}(x^{}_{1\trans},\ldots, x^{}_{k\trans};n,\tau)
-\left\langle
\hat O^{}_{1}(x^{}_{1\trans})\cdots
\hat O^{}_{k}(x^{}_{k\trans})
\right\rangle^{\tau}_{\varrho^{}_{\rm rel}}\right\},
\end{eqnarray}
where $\hat{O}^{}_{i}(x^{}_{i\trans})$ are some Schr\"odinger operators
which may depend on the fermion operators as well as on the EM operators,
and
\begin{equation}
\label{TrCorr}
\widetilde{F}^{}_{1\cdots k}(x^{}_{1\trans},\ldots, x^{}_{k\trans};n,\tau)=
\left\langle
\hat O^{}_{1}(x^{}_{1\trans})\cdots
\hat O^{}_{k}(x^{}_{k\trans})
\right\rangle^{\tau}_{\tilde{\varrho}}
\end{equation}
are the ``equal-time" correlation functions in which the condensate
mode of the EM field is eliminated. Since
the relevant statistical operator~(\ref{RelStOp}) admits Wick's
decomposition, the last term in Eq.~(\ref{TrCorr:Eq}) ensures the
boundary condition of complete weakening of initial correlations.
Note that the explicit knowledge of the statistical operator
$\widetilde{\varrho}(n,\tau)$ is not needed when considering the hierarchy for the
correlation functions.
Use of some truncation procedure is a standard practice in this case.
The hierarchy for correlation functions
will be discussed in subsequent
papers in the context of the derivation of collision integrals.

Another method of handling Eq.~(\ref{ZubEq}) is by considering its formal
solution
 \begin{equation}
 \label{StOp1}
\widetilde{\varrho}(n,\tau)=
\varepsilon \int\limits^{\tau}_{-\infty} d\tau'\,
{\rm e}^{-\varepsilon(\tau -\tau')}\,
{\mathcal U}(\tau,\tau')\,{\varrho}^{}_{\rm rel}(n,\tau')\,
{\mathcal U}^{\dagger}(\tau,\tau'),
 \end{equation}
where the evolution operator can be written as the ordered exponent
 \begin{equation}
 \label{EffEvOp}
{\mathcal U}(\tau,\tau') =
T^{}_{\tau}\,\exp\left\{-i \int\limits^{\tau}_{\tau'}
\hat{\mathcal H}^{\bar\tau}(n)\,d\bar{\tau}  \right\}.
 \end{equation}
After partial integration, the expression~(\ref{StOp1}) becomes
 \begin{equation}
 \label{StOp2}
\widetilde{\varrho}(n,\tau)=
\varrho^{}_{\rm rel}(n,\tau) +
\Delta{\varrho}(n,\tau),
\end{equation}
where
\begin{eqnarray}
\label{DeltaRho}
& &
\hspace*{-20pt}
\Delta{\varrho}(n,\tau)=
-\int\limits^{\tau}_{-\infty} d\tau'\,
{\rm e}^{-\varepsilon(\tau -\tau')}
\nonumber\\[8pt]
& &
\hspace*{40pt}
{\times}\,
{\mathcal U}(\tau,\tau')
\left\{
\frac{\partial {\varrho}^{}_{\rm rel}(n,\tau')}{\partial\tau'}
-i\left[{\varrho}^{}_{\rm rel}(n,\tau'),
\hat{\mathcal H}^{\tau'}(n)\right]
\right\}
{\mathcal U}^{\dagger}(\tau,\tau').
\end{eqnarray}
The representation~(\ref{StOp2}) for the statistical operator allows to
separate the mean-field terms and the collision terms
in Eqs.~(\ref{FDens:EqMot}) and~(\ref{PDens:EqMot}). Taking into account
the self-consistency conditions~(\ref{SelfCons}) and the fact that the
Hamiltonian~(\ref{ZeroHam}) is bilinear in the fermion and photon
operators, we arrive at the equations
 \begin{eqnarray}
 \label{KinEq:F}
& &
\hspace*{-10pt}
\frac{\partial}{\partial\tau}\,
\opdm^{}_{aa'}(x^{}_{\trans},x^{\prime}_{\trans};\tau)=
-i\left\langle \big[
\hat \opdm^{}_{aa'}(x^{}_{\trans},x^{\prime}_{\trans}),
\hat{\mathcal H}^{\tau}_{0}(n)\big]
\right\rangle^{\tau}_{\varrho^{}_{\rm rel}} +
I^{(f)}_{aa'}(x^{}_{\trans},x^{\prime}_{\trans};\tau),
\\[8pt]
\label{KinEq:Ph}
& &
\hspace*{-10pt}
\frac{\partial}{\partial\tau}\,
N^{}_{ll'}(x^{}_{\trans},x^{\prime}_{\trans};\tau)=
-i\left\langle
\big[
\hat{N}^{}_{ll'} (x^{}_{\trans},x^{\prime}_{\trans}),
\hat{H}^{}_{EM}\big]
\right\rangle^{\tau}_{\varrho^{}_{\rm rel}} +
I^{(ph)}_{ll'}(x^{}_{\trans},x^{\prime}_{\trans};\tau),
 \end{eqnarray}
where the collision integrals for fermions and photons are given by
 \begin{eqnarray}
 \label{ColInt:F}
& &
I^{(f)}_{aa'}(x^{}_{\trans},x^{\prime}_{\trans};\tau)=
-i\left\langle \big[
\hat \opdm^{}_{aa'}(x^{}_{\trans},x^{\prime}_{\trans}),
\hat{\mathcal H}^{\tau}_{\rm int}(n)\big]
\right\rangle^{\tau}_{\varrho^{}_{\rm rel}}
\nonumber\\[6pt]
& &
\hspace*{85pt}
{}-i\,{\rm Tr}
\left\{
 \big[
\hat \opdm^{}_{aa'}(x^{}_{\trans},x^{\prime}_{\trans}),
\hat{\mathcal H}^{\tau}_{\rm int}(n)
\big]\,\Delta{\varrho}(n,\tau)\right\},
\\[8pt]
 \label{ColInt:Ph}
& &
I^{(ph)}_{ll'}(x^{}_{\trans},x^{\prime}_{\trans};\tau)=
-i\,{\rm Tr}
\left\{
 \big[
\hat N^{}_{ll'}(x^{}_{\trans},x^{\prime}_{\trans}), \hat{\mathcal
H}^{\tau}_{\rm int}(n) \big]\,\Delta{\varrho}(n,\tau)\right\}.
 \end{eqnarray}
In the presence of a strong EM field, the evolution of the fermion subsystem
is governed predominantly by its interaction with the mean EM field.
Thus, the covariant mean-field kinetic equation for the Wigner
function~(\ref{Wigner:Def}) can be derived from Eq.~(\ref{KinEq:F})
neglecting the collision integral.
This kinetic equation as well as the collision integrals~(\ref{ColInt:F})
and~(\ref{ColInt:Ph}) will be considered in subsequent papers.

\section{Concluding remarks}
We have shown that the hyperplane formalism can serve as the basis
for kinetic theory of QED plasmas in the presence of a strong external field.
The formalism has the advantage that it is manifestly
covariant and therefore allows to introduce different approximations in
covariant form.
Only minor changes with respect to the non-relativistic density
matrix method are introduced, so that many well-developed
approaches can be directly applied to QED plasmas.
For instance, the explicit construction of the statistical operator allows to
incorporate many-particle correlations through the extension of the
set of basic state parameters (see, e.g.,\cite{ZubMorRoep1}). Note also that,
using the Heisenberg picture on hyperplanes,
nonequilibrium Green's
functions can be introduced with respect
to the invariant time parameter $\tau$.
In such a way, the spectral properties
of microscopic dynamics can be incorporated.

The scheme outlined in this paper is also applicable to other field
theories, like QCD transport theory. In QCD, however, additional
problems arise due to its non-Abelian structure, which needs further
considerations. 

Finally, we would like to emphasize once again two key problems
in a covariant density matrix approach to relativistic kinetic theory
in the presence of a strong mean field.
First, it is necessary to perform canonical quantization of the system
on a hyperplane in Minkowski space.
Second, the condensate mode
must be separated from the quantum degrees of freedom at any time.
We have shown how these problems can be solved in the context of QED
plasmas. As a result, a general form of kinetic equations for fermions
and photons was given.

The scheme outlined in this paper is also applicable to some quantum
field models used in QCD transport theory. In this case the non-Abelian
algebra must be worked out to describe the quark-gluon plasma.

\renewcommand{\theequation}{A.\arabic{equation}}
\setcounter{equation}{0}
\section*{Appendix A}
\subsection*{Commutation relations for electromagnetic  field on hyperplanes}
Let us write the constraint equations for the canonical variables
$A^{\mu}_{\trans}$ and $\Pi^{\mu}_{\trans}$  in the form
$\chi^{}_{N}(x^{}_{\trans})=0$, where
\begin{equation}
\label{ConstrFunc}
\begin{array}{ll}
\chi^{}_{1}(x^{}_{\trans})=
\nabla^{}_{\mu} A^{\mu}_{\trans}(x^{}_{\trans}),
&
\qquad
\chi^{}_{2}(x^{}_{\trans})=
\nabla^{}_{\mu} \Pi^{\mu}_{\trans}(x^{}_{\trans}),
\\[6pt]
\chi^{}_{3}(x^{}_{\trans})=
n^{}_{\mu} A^{\mu}_{\trans}(x^{}_{\trans}),
&
\qquad
\chi^{}_{4}(x^{}_{\trans})=
n^{}_{\mu}  {\Pi}^{\mu}_{\trans}(x^{}_{\trans}).
\end{array}
\end{equation}

For any functionals $\Phi^{}_{1}$ and $\Phi^{}_{2}$
of the field variables $A^{}_{\trans}$ and $\Pi^{}_{\trans}$,
we define the Poisson bracket
\begin{equation}
\label{P-brack}
\hspace*{-15pt}
\left[\Phi^{}_{1},\Phi^{}_{2}\right]^{}_{\rm P}\equiv
\int^{}_{\sigma^{}_{n,\tau}} d\sigma
\left\{
\frac{\delta\Phi^{}_{1}}
{\delta A^{\mu}_{\trans}(x^{}_{\trans})}\,
\frac{\delta\Phi^{}_{2}}
{\delta \Pi^{}_{\trans\mu}(x^{}_{\trans})}
-\frac{\delta\Phi^{}_{2}}
{\delta A^{\mu}_{\trans}(x^{}_{\trans})}\,
\frac{\delta\Phi^{}_{1}}
{\delta \Pi^{}_{\trans\mu}(x^{}_{\trans})}
\right\},
\end{equation}
where the constraints are ignored in calculating the functional
derivatives. Applying this formula to the canonical variables we obtain
\begin{equation}
\label{APi:P-br}
\left[
A^{\mu}_{\trans}(x^{}_{\trans}),
\Pi^{}_{\trans\nu}(x^{\prime}_{\trans})
\right]^{}_{\rm P}=\delta^{\mu}_{\ \nu}\,
\delta^{3}(x^{}_{\trans}-x^{\prime}_{\trans})
\end{equation}
with the three-dimensional delta function~(\ref{DeltaFunc}).
All other Poisson brackets for the canonical variables are equal to
zero. In the Dirac terminology,  functions~(\ref{ConstrFunc})
correspond to {\em second class\/} constraints since the matrix
\begin{equation}
\label{Cmatr}
C^{}_{NN'}(x^{}_{\trans}, x^{\prime}_{\trans})=
\left[
\chi^{}_{N}(x^{}_{\trans}),\chi^{}_{N'}(x^{\prime}_{\trans})
\right]^{}_{\rm P}
\end{equation}
is non-singular. A straightforward calculation of the Poisson brackets
shows that the non-zero elements of $C$ are
\begin{eqnarray}
\label{Cmatr:el}
& &
C^{}_{12}(x^{}_{\trans}, x^{\prime}_{\trans})=
- C^{}_{21}(x^{}_{\trans}, x^{\prime}_{\trans})=
-\nabla^{}_{\mu} \nabla^{\mu}
\delta^3(x^{}_{\trans}-x^{\prime}_{\trans}),
\nonumber\\[6pt]
& &
C^{}_{34}(x^{}_{\trans}, x^{\prime}_{\trans})=
- C^{}_{43}(x^{}_{\trans}, x^{\prime}_{\trans})=
\delta^3(x^{}_{\trans}-x^{\prime}_{\trans}).
\end{eqnarray}
According to the general quantization scheme~\cite{Dirac50,Weinberg96},
commutation relations for canonical operators
are defined by the Dirac brackets for classical canonical
variables. In our case  the Dirac brackets are written as
\begin{eqnarray}
\label{DiracBr:Def}
& &
\hspace*{-10pt}
\left[\Phi^{}_{1},\Phi^{}_{2}\right]^{}_{\rm D}=
\left[\Phi^{}_{1},\Phi^{}_{2}\right]^{}_{\rm P}
\nonumber\\[8pt]
& &
\hspace*{30pt}
{}- \int_{\sigma^{}_{n,\tau}} d\sigma
\int_{\sigma^{}_{n,\tau}} d\sigma'
\left[\Phi^{}_{1},\chi^{}_{N}(x^{}_{\trans})\right]^{}_{\rm P}
C^{-1}_{NN'}(x^{}_{\trans},x^{\prime}_{\trans})
\left[\chi^{}_{N'}(x^{\prime}_{\trans}),
\Phi^{}_{2}\right]^{}_{\rm P}
\end{eqnarray}
(summation over repeated indices). The inverse matrix,
$C^{-1}_{NN'}(x^{}_{\trans},x^{\prime}_{\trans})$, satisfies the
equation
\begin{equation}
\label{InverC:Eq}
\int_{\sigma^{}_{n,\tau}} d\sigma''\,
C^{}_{NN''}(x^{}_{\trans},x^{\prime\prime}_{\trans})\,
C^{-1}_{N''N'}(x^{\prime\prime}_{\trans},x^{\prime}_{\trans})=
\delta^{}_{NN'}\,\delta^{3}(x^{}_{\trans}-x^{\prime}_{\trans}).
\end{equation}
Since the matrix elements~(\ref{Cmatr:el}) of $C$ depend on the
difference $x^{}_{\trans}-x^{\prime}_{\trans}$, Eq.~(\ref{InverC:Eq})
can  be solved for $C^{-1}$ using a Fourier transform on
$\sigma^{}_{n,\tau}$, which is defined for any function $f(x)$ as
\begin{equation}
\label{FourTrans}
\tilde{f}(\tau,p^{}_{\trans})=
\int d^4x\, {\rm e}^{ip\cdot x}\, \delta(x\cdot n -\tau)\,f(x).
\end{equation}
The inverse transform is
\begin{equation}
\label{FourTrans:Inv}
f(x)\equiv f(\tau,x^{}_{\trans})=
\int \frac{d^{4}p}{(2\pi)^{3}}\,
{\rm e}^{-ip\cdot x}\, \delta(p\cdot n) \tilde f(\tau,p^{}_{\trans}).
\end{equation}
If we perform the Fourier transformation in Eq.~(\ref{InverC:Eq}), we
find by inserting~(\ref{Cmatr:el}) that  the non-zero
elements of $C^{-1}$ are
\begin{eqnarray}
\label{C-1:el}
& &
C^{-1}_{12}(x^{}_{\trans},x^{\prime}_{\trans})=
- C^{-1}_{21}(x^{}_{\trans},x^{\prime}_{\trans})=
- \int \frac{d^{4}p}{(2\pi)^{3}}\,
{\rm e}^{-ip\cdot (x -x')}\, \delta(p\cdot n) \frac{1}{p^{2}_{\trans}},
\nonumber\\[6pt]
& & C^{-1}_{34}(x^{}_{\trans},x^{\prime}_{\trans})= -
C^{-1}_{43}(x^{}_{\trans},x^{\prime}_{\trans})= -
\delta^{3}(x^{}_{\trans}-x^{\prime}_{\trans}).
\end{eqnarray}
Now the Dirac brackets~(\ref{DiracBr:Def}) for the canonical variables
are easily calculated and we obtain
\begin{eqnarray}
\label{DiracBr:Can1}
& & \left[
A^{\mu}_{\trans}(x^{}_{\trans}), {\Pi}^{\nu}_{\trans}(x^{\prime}_{\trans})
\right]^{}_{\rm D}= c^{\mu\nu}(x^{}_{\trans} -x^{\prime}_{\trans}),
\\[6pt]
& &
\label{DiracBr:Can2}
\left[A^{\mu}_{\trans}(x^{}_{\trans}),
{A}^{\nu}_{\trans}(x^{\prime}_{\trans})
\right]^{}_{\rm D}=\left[{\Pi}^{\mu}_{\trans}(x^{}_{\trans}),
{\Pi}^{\nu}_{\trans}(x^{\prime}_{\trans})
\right]^{}_{\rm D}=0,
\end{eqnarray}
where the functions $c^{\mu\nu}(x^{}_{\trans} -x^{\prime}_{\trans})$ are given
by Eq.~(\ref{DiracDelta}).
According to the general quantization rules, the commutation relations
for canonical operators correspond to $i[\ldots]^{}_{\rm D}$.
Thus, in the hyperplane formalism, the commutation relations for the
operators of EM field are given by~(\ref{Comm:Can}) and~(\ref{Comm:CanZero}).
Obviously these relations are valid in the
Schr\"odinger and Heisenberg pictures.

\renewcommand{\theequation}{B.\arabic{equation}}
\setcounter{equation}{0}
\section*{Appendix B}
\subsection*{Anticommutation relations for the Dirac field on hyperplanes}
To find the anticommutation relations for the fermion
operators  on the hyperplane $\sigma^{}_{n,\tau}$, it is sufficient
to consider a free Dirac field. Our starting point is the standard
quantization scheme in the frame where
$x^{\mu}=(t,\vek{r})$ and $n^{\mu}=(1,0,0,0)$
(see, e.g.,~\cite{Greiner96}).
In that case the field  operators $\hat{\psi}^{}_{a}$ and
$\,\hat{\!\bar\psi}^{}_{a}$
can be written in terms of creation and
annihilation operators according to
$$
\begin{array}{l}
\displaystyle
\hat{\psi}_{a}^{}(x) = \int \frac{d^4 p}{(2\pi )_{}^{3/2}} \;
\frac{\delta (p_{}^{0}
- \epsilon (\vek{p}))}{\sqrt{2 \epsilon (\vek{p})}} \sum_{s=\pm 1}
\left[
\hat{b}_{s}^{}(p) u_{as}^{}(p) \e_{}^{-ip\cdot x}
+ \hat{d}_{s}^{\dagger}(p) v_{as}^{}(p)\e_{}^{ip\cdot x}
\right],
\\[18pt]
\displaystyle
\,\hat{\!\bar\psi}_{\!a}^{}(x) = \int \frac{d^4 p}{(2\pi )_{}^{3/2}} \;
\frac{\delta (p_{}^{0}
- \epsilon (\vek{p}))}{\sqrt{2 \epsilon (\vek{p})}} \sum_{s=\pm 1}
\left[
\hat{d}_{s}^{}(p) \bar v_{as}^{}(p) \e_{}^{-ip\cdot x}
+ \hat{b}_{s}^{\dagger}(p) \bar u_{as}^{}(p) \e_{}^{ip\cdot x}
\right],
\end{array}
$$
where $\epsilon (\vek{p}) =\sqrt{\vek{p}_{}^{2} + m_{}^{2}}$
is the free fermion dispersion relation.
Constructing the
expression
$\{\hat{\psi}_{a}^{}(x),\,\hat{\!\bar\psi}_{\!a'}^{}(x')\}$
for two arbitrary space-time points and recalling
the anticommutation relations
\begin{equation}
\label{Acommbd}
\left\{
\hat{b}_{s}^{}(\vek{p}) , \hat{b}_{s'}^{\dagger}(\vek{p}')
\right\}
=
\left\{
\hat{d}_{s}^{}(\vek{p}) , \hat{d}_{s'}^{\dagger}(\vek{p}')
\right\}
= \delta_{ss'} \delta_{}^{3}(\vek{p} - \vek{p}'),
\end{equation}
as well as polarization sums
$$
\sum_{s=\pm 1} u^{}_{as}(p)\bar{u}^{}_{a's}(p)=
\left[\gamma^{\mu}p^{}_{\mu} +m \right]^{}_{aa'},
\qquad
\sum_{s=\pm 1} v^{}_{as}(p)\bar{v}^{}_{a's}(p)=
\left[\gamma^{\mu}p^{}_{\mu} - m \right]^{}_{aa'},
$$
we arrive at
\begin{eqnarray}
\label{Tmp1}
\left\{
\hat{\psi}_{a}^{}(x),\,\hat{\!\bar\psi}_{\!a'}^{}(x')
\right\}
&=& \int \frac{d^3\vek{p}}{(2\pi )_{}^{3}} \;
\frac{1}{2\epsilon (\vek{p})}
\left\{
\left[\gamma^{\mu} p^{}_{\mu}
 +  m \right]_{aa'}
\e^{-ip\cdot (x-x')}
\right.
\nonumber\\[6pt]
& &
\hspace*{60pt}
\left.
+ \left[ \gamma^\mu p^{}_\mu -  m \right]_{aa'}
\e^{ip\cdot (x-x')} \right\},
\end{eqnarray}
where $p^{0} = \sqrt{\vek{p}_{}^{2} + m^2}$.
Using
\begin{equation}
\label{Tmp3}
\int \frac{d^3 \vek{p}}{(2 \pi )^3} \;
\frac{1}{2 \epsilon (\vek{p})}
=
\left.
\int \frac{d^4 p}{(2 \pi )^3}\;
\delta (p^2 - m^2) \right|_{p^{0} > 0}\, ,
\end{equation}
Eq.~(\ref{Tmp1}) can be rewritten in a Lorentz invariant form
\begin{eqnarray}
\label{Tmp4}
& &
\hspace*{-25pt}
\left\{
\hat{\psi}_{a}^{}(x),\,\hat{\!\bar\psi}_{a'}^{}(x')
\right\}
=
\int \frac{d^4 p}{(2 \pi )^3}\;
\left\{
\left[\gamma^\mu  p^{}_\mu +  m \right]_{aa'}
\e^{-ip\cdot (x-x')}
\left. \delta (p^2 - m^2)\right|_{p^0 > 0}
\right.
\nonumber\\[6pt]
& &
\hspace*{80pt}
\left.
+ \left[ \gamma^\mu  p^{}_\mu -  m \right]_{aa'}
\e^{ip\cdot (x-x')}
\left. \delta (p^2 - m^2)\right|_{p^0 > 0}
\right\}.
\end{eqnarray}
The  anticommutation relation on the hyperplane $\sigma^{}_{n,\tau}$
is now obtained by setting $x=n\tau +x^{}_{\trans}$ and
$x'=n\tau +x^{\prime}_{\trans}$. In calculating the integrals, it is
convenient to use the decomposition
$p^{\mu}= n^{\mu} p^{}_{\longi} + p^{\mu}_{\trans}$,
($p^{}_{\longi}>0$).
Then we get
\begin{eqnarray}
\label{Tmp5}
& &
\hspace*{-20pt}
\left\{
\hat{\psi}_{a}^{}(\tau ,x_{\trans}^{}),
\,\hat{\!\bar\psi}_{\!a'}^{}(\tau ,x_{\trans}^{\prime} )
\right\}
= \int \frac{d^4 p}{(2\pi )^3}\,
\frac{\delta (p_{\longi}^{ } - \epsilon (p_{\trans}^{}))}
{2\epsilon (p_{\trans}^{})}
\nonumber\\[8pt]
& &
\hspace*{60pt}
{}\times\left\{
\left[
\gamma_{\longi}^{} p_{\longi}^{}
+ \gamma^{\mu}_{\trans} p^{}_{\trans\mu} +  m
\right]_{aa'}^{}
\e^{-ip_{\trans\mu}^{} (x_{\trans}^{\mu} - x_{\trans}^{\prime\mu})}
\right.
\nonumber\\[6pt]
& &
\hspace*{120pt}
\left.
{}+
\left[
\gamma_{\longi}^{}p_{\longi}^{}
+ \gamma^{\mu}_{\trans}  p^{}_{\trans\mu} -  m
\right]_{aa'}^{}
\e^{ip_{\trans \mu}^{}(x_{\trans}^{\mu} - x_{\trans}^{\prime\mu})}
\right\}
\end{eqnarray}
with the dispersion relation on the hyperplane
\begin{equation}
\label{Disp}
\epsilon (p_{\trans}^{})
= \sqrt{-p_{\trans \mu}^{} p_{\trans}^{\mu} + m^{2}}.
\end{equation}
Finally, changing  the variable
$p_{\trans} \rightarrow - p_{\trans}$ in the second integral in
Eq.~(\ref{Tmp5}), we obtain the anticommutation
relation~(\ref{Anticomm1}). The relations~(\ref{Anticomm2}) can be derived
by the same procedure.

\section{Acknowledgments}
The main part of this work was conducted during visits in Rostock and
Moscow. V.M.~Morozov would like to thank the ``Deutsche
Forschungsgemeinschaft'' and A.~H\"oll the ``Studienstiftung des deutschen 
Volkes'' and the ``Deutsche Forschungsgemeinschaft'' for supporting
this work.

\end{document}